\documentclass[11pt]{article}
\usepackage{graphics}
\usepackage{amssymb}
\usepackage{psfrag}
\usepackage{epsfig}
\usepackage{psfig}
\usepackage{epsf}
\usepackage{float}
\usepackage{latexsym}
\usepackage[german,USenglish]{babel}
\newcommand{\vs}{\vspace{-0.2cm}}
\newcommand{\beq}{\begin{equation}}
\newcommand{\eeq}{\end{equation}}
\newcommand{\beqa}{\begin{eqnarray}}
\newcommand{\eeqa}{\end{eqnarray}}
\newcommand{\nn}{\nonumber \\ }
\newcommand{\bma}{\begin{array}{cc}}
\newcommand{\ema}{\end{array}}

\newcommand{\krig}[1]{\stackrel{\circ}{#1}}

\def\3{{\ss}}

\def\vek #1 {\overrightarrow {#1}}
\newcommand{\fet}[1]{\mbox{\boldmath $#1$}}

\begin{document}

\hfill {\tiny FZJ-IKP(TH)-2002-14}

\vspace{1cm}

\begin{center}

{{\Large\bf Nuclear forces in the chiral limit}}

\end{center}

\vspace{.3in}

\begin{center}

{\large 
E. Epelbaum,$^\dagger$\footnote{email: 
                           evgeni.epelbaum@tp2.ruhr-uni-bochum.de}
Ulf-G. Mei{\ss}ner},$^\ddagger$\footnote{email: 
                           u.meissner@fz-juelich.de}
W. Gl\"ockle$^\dagger$\footnote{email:
                           walter.gloeckle@tp2.ruhr-uni-bochum.de}

\bigskip

$^\dagger${\it Ruhr-Universit\"at Bochum, Institut f{\"u}r
  Theoretische Physik II,\\ D-44870 Bochum, Germany}\\

\bigskip

$^\ddagger${\it Forschungszentrum J\"ulich, Institut f\"ur Kernphysik 
(Theorie),\\ D-52425 J\"ulich, Germany} \\
 and       \\
{\it Karl-Franzens-Universit\"at Graz, Institut f\"ur Theoretische Physik\\ 
A-8010  Graz, Austria}

\end{center}

\vspace{.3in}

\thispagestyle{empty}

\begin{abstract}
\noindent We investigate the behaviour of the nuclear forces as a function of the
light quark masses (or, equivalently, pion mass) 
in the framework of chiral effective field theory at next-to-leading
order. The nucleon--nucleon force is described in terms of one and two--pion exchange
and local short distance operators, which depend explicitly and implicitly on the
quark masses. The pion propagator becomes Coulomb-like in the chiral limit and thus
one has significant scattering in all partial waves. The pion-nucleon coupling
depends implicitly on the quark masses and we find that it becomes stronger in the
chiral limit. There is a further quark mass dependence in the S-wave four--nucleon
couplings, which can be estimated by means of dimensional analysis. We find that
nuclear physics in the chiral limit becomes natural. There are no new bound states,
the deuteron binding energy is $B_D^{\rm CL} = 9.6 \pm 1.9  
{ {+ 1.8} \atop {-1.0}} \,$MeV, and the S--wave scattering
lengths take values of a few fm, $a_{\rm CL} (^1S_0) = -4.1 \pm 1.6
{{+ 0.0} \atop 
{-0.4}} \,$fm and 
$a_{\rm CL} (^3S_1) = 1.5 \pm 0.4
{{+ 0.2} \atop \\ 
{ -0.3}}\,$fm. We also discuss the extrapolation to larger pion
masses pertinent for the extraction of these quantities from lattice simulations.
\end{abstract}

\noindent
\vfill

\pagebreak
\section{Introduction}
\def\theequation{\arabic{section}.\arabic{equation}}
\setcounter{equation}{0}
\label{sec:intro}

The application of (chiral) effective field theory (EFT) 
to few--nucleon systems has become a 
subject of vigorous research interest in the last years. One successful method
is based upon the original idea of Weinberg \cite{wein}
to apply the standard methods of Chiral Perturbation Theory (CHPT) to the kernel of the 
corresponding integral equation (Lippmann--Schwinger equation in case of two nucleons and
Faddeev--Yakubovsky equations for three and more nucleons), and consequently solving it numerically.
The necessity of summing the interactions to an infinite order is caused by  
the nonperturbative aspects  of the problem (shallow nuclear bound states, 
unnaturally large S--wave scattering lengths), quite in contrast to the purely
perturbative treatment possible in the pion and pion--nucleon sectors.  
The complete analysis of two-- and three--nucleon systems and, in addition, 
of the $\alpha$--particle has been recently performed 
up to  next--to--next--to--leading order (NNLO) in the chiral expansion 
\cite{EGM1,EGM2,3Nno3NF,new3N}. Going to NNLO turns out to be sufficient to describe 
accurately most of the low--energy observables.

CHPT is an effective field theory based upon the approximate and 
spontaneously broken chiral symmetry of QCD and allows for a model--independent
and systematic way of calculating low--energy hadronic observables. This is 
achieved by a simultaneous expansion in small external momenta and quark (pion)
masses around the chiral limit, in which the light quarks 
(and, as a consequence, also Goldstone
bosons) are massless. Since this expansion is well defined for sufficiently small
quark masses ($m_q$), hadronic properties 
at low energies are expected  not to change strongly in the limit $m_q \rightarrow 0$.
This feature is of  crucial importance for the applicability of CHPT
and is certainly valid in the purely pionic sector. In the chiral limit the 
interaction between pions becomes arbitrarily weak for vanishing external momenta. 
The complete one--loop analysis for the two flavor case 
has been performed by Gasser and Leutwyler in 1984 \cite{GL84}. By now, 
the two--loop analysis of the $\pi\pi$  scattering  and other selected
S--matrix elements and transition currents has become available, see e.g. \cite{twoloop}.
In the pion--nucleon case the situation is  similar,
since the coupling of the Goldstone bosons to matter fields vanishes at 
small momentum transfer in the chiral limit as well.
The loop expansion in the presence of the additional mass scale (i.e. the nucleon
mass in the chiral limit) is somewhat more tricky, but various methods have
been invented to allow for a consistent and symmetry preserving power counting. 
The purpose of the present work is to consider the 
nucleon--nucleon (NN) interaction in the chiral limit. 
Clearly, this case requires additional care because 
of the nonperturbative aspects of the problem and also due to the fact
that the interaction between nucleons does not become weak in the  
chiral limit.\footnote{Moreover, it even gets stronger in most channels, 
as we will show below.} In principle, at very low energies one may consider a
theory of nucleons only interacting via contact interactions, the so--called
pionless EFT (see e.g. \cite{sanspion} and references therein). 
While such a scheme has been shown to be a precise calculational
tool, the link to the chiral symmetry is lost and we thus eschew such type of EFT 
here.\footnote{In the pionless EFT the pion mass is considered to be a large 
scale and the limit $M_\pi \rightarrow 0$ cannot be performed.}
We stress that the question about the $M_\pi$--dependence of nuclear forces 
is not only of academic interest, but also of
practical use since such studies of quark mass dependences can be used to 
interpolate lattice gauge theory data, which are usually obtained for quark (pion) 
masses much larger than physical ones. Furthermore, the deuteron and other light
nuclei can be considered as laboratories to test the recent ideas of the time variation
of certain fundamental couplings constants (see e.g. \cite{uzan} and references therein).

The first step in extrapolating the properties of the NN system for 
vanishing quark mass (exact chiral symmetry of QCD) was done
by Bulgac et al. \cite{Bul97} a
few years ago. They stressed that the vanishing pion mass may potentially
lead to strong effects in the nucleon--nucleon P-- and higher partial waves.  
One could even not exclude a priori the presence of bound states in some 
of the P--waves in the chiral limit. This indicates that in case of few--nucleon systems,
the expansion around  the chiral limit might be complicated or even not well defined.
Based upon the one--pion exchange potential (OPEP)
and taking into account only its explicit dependence on the pion mass, the authors 
of ref.~\cite{Bul97} came to the conclusion that no bound states appear in P-- and 
higher partial waves in the chiral limit. Performing first--order perturbation 
theory  and using the deuteron wave--functions from some
phenomenological potential models, Bulgac et al. 
found a reduction of about 1 MeV (that is by 50\%) in the deuteron binding energy. 
The main conclusion of ref.~\cite{Bul97} was therefore that 
``$\ldots$ physics of nuclei in a universe 
where $M_\pi=0$ would be similar to what is actually observed.''
However, the interpretation of the results 
found in ~\cite{Bul97} requires some caution not only due to the fact that 
first--order perturbative results can not be trusted in that 
case,\footnote{As pointed out in \cite{Beane01},
one gets a deuteron binding energy $B_d\sim 4.1$ MeV 
if the pion mass is set to zero in the 
AV18 potential. Thus the effect is of the opposite sign to the one found 
in \cite{Bul97} using  first--order perturbation theory.} 
but also because only the explicit dependence on the pion mass has been taken 
into account. In the language of  chiral effective field theory,
taking into account only the explicit $M_\pi$--dependence 
in the OPEP corresponds (roughly) to the LO approximation, 
which is not accurate enough to lead to any quantitative conclusions.

More recently, Beane et al. re-analyzed the situation in the 
$^1S_0$ and $^3S_1-{^3D}_1$ channels 
including also some implicit dependence 
of the NN interaction on the pion mass \cite{Beane01}. 
They considered the OPEP as an approximation
to the NN interaction for distances larger than some matching radius $R_\star$. 
For shorter distances the potential was approximated by a square well, 
which can be viewed as a smeared out delta--function 
counter term. In the $^3S_1$--$^3D_1$--channel they also included an 
additional energy--dependent  contribution in the short--range part of the potential.
Adjusting the two parameters related to  the short range part of the interaction, 
Beane et al. performed  exact renormalization of the corresponding Schr\"odinger equation, 
i.e. the calculated low--energy observables do not depend on the matching radius 
$R_\star$. This way of handling the diverging Schr\"odinger equation 
is closely related to their previous work \cite{BeaneLimC}. 
The situation in the chiral limit has been analyzed in two ways: 
(i)
Keeping only the explicit $M_\pi$ dependence of the OPEP an increase of the deuteron 
binding energy by about a factor of two has been found, $B_d \sim 4.2$ MeV.
(ii)
The authors of ref.\cite{Beane01} also made an attempt to take into account the implicit 
$M_\pi$--dependence, which changes the strength of the OPEP in the chiral limit. 
They further assumed that these corrections can be represented by the leading 
logarithmic terms in the chiral expansion.\footnote{Note that this assumption goes  
beyond standard CHPT and is not applicable in a general case. 
We will come back to this point later on.}
Quite surprisingly, the results for deuteron binding energy were found to 
change completely after taking into account the implicit $M_\pi$--dependence 
in this manner. It turned out that deuteron becomes unbound for the pion mass
smaller than $\sim 100$ MeV. 
Another interesting issue discussed in \cite{Beane01} refers to the 
possibility of matching the results 
obtained in chiral effective field theory with the ones from lattice QCD \cite{Fukug95}, 
which requires extrapolation away from the physical value of $M_\pi$. 

While this paper was in preparation, an NLO analysis of the quark mass dependence in the 
$^1S_0$ and $^3S_1$--$^3D_1$ channels by Beane and Savage based upon the power counting 
introduced in ref.~\cite{Beane01} has become available \cite{BeaneNEW},
improving and extending the results of ref.~\cite{Beane01}. Apart from the 
OPE and short--range terms in the effective potential, the leading TPE contribution 
(in the chiral limit) has been taken into account. The authors found evidence for nearly 
all possible scenarios such as a bound or unbound deuteron and di--neutron in the $^1S_0$ 
channel and could not make any predictions for the corresponding scattering lengths
due to the lack of knowledge of the low--energy constants related to contact terms with 
one insertion proportional to $M_\pi^2$. 
We will come back to this work and to comparison to our results in the 
next-to-last section.

The aim of this work is to perform a  complete NLO calculation
in the framework of a modified Weinberg power counting and to try to clarify the 
presently somewhat controversial situation about the chiral limit of 
the nucleon--nucleon interaction \footnote{This is similar to what was done in
\cite{BeaneNEW} but differs markedly in some aspects from that work as will be
shown below.}. To do that, 
we have to consider apart from the OPEP and 
contact interactions also the leading two--pion exchange
potential (TPEP). We will discuss in detail the renormalization of the 
NN potential at NLO since this is of utmost importance for such type of analysis. 
We will not use of the assumption of the dominance of chiral logarithms but
rather work with the complete CHPT expressions at this order. The corresponding
low--energy constants (LECs) will be taken from the analysis of various processes.
We will demonstrate that using only the leading logarithms
leads to  results for the renormalization of the axial coupling $g_A$ 
incompatible with the present analyses
of the $\pi N$ system and of the process $\pi N \to \pi \pi N$. We also perform an 
extrapolation of the $^1S_0$ and $^3S_1$ scattering lengths  
and deuteron binding energy $B_{\rm D}$  for the values of the pion mass in the 
range  $0 < M_\pi < 300$ MeV.

Our manuscript is organized as follows. In section~\ref{sec:pot} we consider the
chiral effective nucleon-nucleon potential at next--to--leading order, with particular
emphasis on the aspects of renormalization pertinent to derive the desired quark mass
dependences. Such an analysis is not yet available in the literature. We also demonstrate
the equivalence between the S--matrix approach used to analyze the
pion--nucleon sector  and the non--covariant projection formalism
employed to construct the NN potential. We systematically work out all
explicit and implicit quark mass dependences of the various contributions due to
one-- and two--pion exchanges as well as contact interactions. Results for the NN phase
shifts, the deuteron binding energy and the S--wave scattering lengths are collected in
section~\ref{sec:res}. Will also discuss in detail the differences to the earlier
EFT work \cite{Beane01,BeaneNEW}.  A brief summary and further discussion is presented in
section~\ref{sec:summ}.

\section{Nucleon--nucleon potential at next--to--leading order}
\def\theequation{\arabic{section}.\arabic{equation}}
\setcounter{equation}{0}
\label{sec:pot}

In this section, we remind the reader on the structure of the NN potential at NLO and 
perform its complete renormalization, which is of crucial importance for considering 
the chiral limit. This has not been done in our 
previous work \cite{EGM2}, since we were only interested in 
constructing the potential for the physically relevant case.  
Since different methods have been applied in the literature 
to define the effective interaction it appears necessary to 
begin our considerations with 
a brief overview of some commonly used formalisms.
This topic is also of technical importance because  the pion--exchange
is directly linked to the pion--nucleon sector,  and thus consistency
with the field theoretical NN potential has to be demonstrated.
The reader more interested in the explicit form of the potential at NLO
exhibiting explicit and implicit pion mass dependences may directly
proceed to section~\ref{sec:NLO}.

\subsection{Construction of the NN potential from field theory}
\def\theequation{\arabic{section}.\arabic{equation}}
%setcounter{equation}{0}
\label{sec:constrpot}

The construction of a potential from field theory is a well known and 
intensively studied problem in nuclear physics. Historically, the  
important conceptual achievements in that direction have been done in the fifties 
in the context of the so called meson field theory. The problem can be formulated in 
the following way: given some field theoretical Lagrangian for interacting mesons 
and nucleons, how can one reduce the (infinite dimensional) equation of motion for 
mesons and nucleons to an effective Schr\"odinger equation for nucleonic 
degrees of freedom, 
which can then be solved by standard methods?
It goes beyond the scope of this paper to discuss the whole variety of the 
different techniques which have been developed to construct an effective interactions, see ref.~\cite{Phil59}
for a comprehensive review.
We will now briefly introduce two methods which will be important for 
our further considerations.
 
The first method is closely related to the field theoretical S--matrix given 
(in the interaction representation) by
\beq
S_{\alpha \beta} 
= \delta_{\alpha \beta} - 2 \pi i \delta ( E_\alpha - E_\beta ) T_{\alpha \beta}\,,
\eeq
where the T--matrix $T_{\alpha \beta}$ satisfies:
\beq
T_{\alpha \beta} =
V_{\alpha \beta} + V_{\alpha \gamma} (E_\beta - E_\gamma 
+ i \epsilon)^{-1} T_{\gamma \beta}\,.
\eeq
Inverting the last equation one obtains the following equation for 
the effective potential $V_{\alpha \beta}$:
\beq
\label{v1}
V_{\alpha \beta} = T_{\alpha \beta} 
- T_{\alpha \gamma} (E_\beta - E_\gamma 
+ i \epsilon)^{-1} T_{\gamma \beta} + \ldots\,,
\eeq
where the ellipsis refer to higher order iterations in the T--matrix.
The effective potential $V_{\alpha \beta}$ can now be obtained 
in terms of a perturbative expansion 
for the T--matrix. For example, in the Yukawa theory one can calculate 
$T_{\alpha \beta}$ (and also $V_{\alpha \beta}$) in terms
of a power series expansion in the squared meson--nucleon coupling constant $g^2$. 
It has to be pointed out
that the potential $V_{\alpha \beta}$ in eq.~(\ref{v1}) is not defined unambiguously 
since only  
on the energy shell T--matrix elements are known and well--defined in field theory. 
In fact, at any fixed order of the perturbative expansion (\ref{v1})
one can perform an arbitrary off--the--energy shell extension of the potential, which 
will then  
affect even on the energy shell matrix elements at higher orders. In addition, one 
has a freedom to 
perform field redefinitions, which changes off shell S--matrix elements and, as a 
consequence, off shell matrix
elements of the potential. Such an ambiguity of the effective potential is a general 
property and 
does not pose any problem, since the latter is not observable. Only observable 
quantities, i.e. on shell
S--matrix elements, binding energies, $\ldots$, are defined unambiguously in 
quantum field theory.

The method introduced above has been used by Kaiser et al. 
to derive the NN potential from the chiral Lagrangian 
\cite{KBW,allKaiser} and will be refered to  in what 
follows as the S--matrix method. 
Its close relation to the scattering amplitude makes it possible to apply the 
standard field theoretical techniques for calculation the effective interactions.

Another well known scheme is often refered to as the 
Tamm--Dancoff method \cite{Ta45,Dan50}. 
The starting point is the time--independent Schr\"odinger equation 
\beq
\label{schroed1}
(H_0 + H_I) | \Psi \rangle = E | \Psi \rangle\,,
\eeq
where $|\Psi \rangle$ denotes an eigenstate of the Hamiltonian $H$ 
with the eigenvalue $E$. 
One can now divide the full Fock space into the nucleonic subspace $|\phi \rangle$ 
and the complementary one $|\psi \rangle$  and rewrite the Schr\"odinger equation 
(\ref{schroed1}) as 
\begin{equation}
\label{schroed2}
\left( \begin{array}{cc} \eta H \eta & \eta H \lambda \\ 
\lambda H \eta & \lambda  H 
\lambda \end{array} \right) \left( \begin{array}{c} | \phi \rangle \\ 
| \psi \rangle \end{array} \right)
= E  \left( \begin{array}{c} | \phi \rangle \\ 
| \psi \rangle \end{array} \right)~,
\quad \,
\end{equation}
where we introduced the projection operators $\eta$ and 
$\lambda$ such that $|\phi \rangle = \eta | \Psi \rangle$,
$| \psi \rangle = \lambda | \Psi \rangle$.
Expressing the state $| \psi \rangle$ from the second line 
of the matrix equation (\ref{schroed2}) as 
\begin{equation}
\label{5.3}
| \psi \rangle = \frac{1}{ E - \lambda H \lambda}  H  | \phi \rangle~,
\end{equation}
and substituting this into
the first line we obtain the Schr\"odinger--like equation for the projected 
state $| \phi \rangle$:
\begin{equation}
\label{TDschroed}
\left( H_0 + V_{{\rm eff}} ( E ) \right) | \phi \rangle  = E | \phi \rangle \,,
\end{equation}
with an effective potential $V_{\rm eff} (E)$ given by
\begin{equation}
\label{TDpot}
V_{\rm eff} (E)= \eta H_I \eta + \eta H_I \lambda 
\frac{1}{E - \lambda H \lambda} \lambda H_I \eta  \,\, .
\end{equation}
Eq.~(\ref{TDschroed}) differs from the ordinary 
Schr\"odinger equation by the fact that 
the effective potential $V_{{\rm eff}} ( E )$ depends explicitly on the energy, 
as it is obvious from 
the definition (\ref{TDpot}). In fact, such an explicit energy dependence 
can be eliminated from the 
potential as discussed e.g. in ref.\cite{Okubo54}. Indeed, 
the effective potential $V_{\rm eff} (E)$ 
up to the fourth order in the coupling constant $g$ in the Yukawa theory 
with $H_I = g H^1$ is given 
in a symbolic form as
\beq
\label{TDg4}
(V_{\rm eff} (E))_{mn} = g^2 \frac{H^1_{m \alpha} \, H^1_{\alpha n}}{E - E_\alpha}
+ g^4 \frac{H^1_{m \alpha}  \, H^1_{\alpha \beta} \,  
H^1_{\beta \gamma} \,  H^1_{\gamma n}}{(E - E_\alpha)
(E - E_\beta)(E - E_\gamma)} + \mathcal{O} (g^6)\,.
\eeq
We denote here the states from the $\eta$--subspace by latin letters and 
the ones from the  $\lambda$--subspace by greek letters.
Using the relation 
\beq
\label{endif}
(E - E_n)\langle n | \phi \rangle = \mathcal{O} (g^2) \langle n | \phi \rangle\,,
\eeq
we can substitute $E$ in the denominator of the last term in eq.~(\ref{TDg4}) by $E_n$.
Thus we can now express eq.~(\ref{TDschroed}) in the form:
\beq
\label{TDgg4}
\left( g^2 \frac{H^1_{m \alpha} \, H^1_{\alpha n}}{E - E_\alpha} 
+ g^4 \frac{H^1_{m \alpha}  \, H^1_{\alpha \beta} \,  
H^1_{\beta \gamma} \,  H^1_{\gamma n}}{(E - E_\alpha)
(E - E_\beta)(E - E_\gamma)} \right) \langle n | \phi \rangle  + \mathcal{O} (g^6) =
\left( E - E_m \right) \langle m | \phi \rangle \,.
\eeq
The first term in the left--hand side of this equation can be expressed in the form
\beq
\label{pr1}
g^2 \frac{H^1_{m \alpha} \,  H^1_{\alpha n}}{E - E_\alpha}  \langle n | \phi \rangle = \left( g^2 
\frac{H^1_{m \alpha}  \, H^1_{\alpha n}}{E_n - E_\alpha} + g^2 
\frac{H^1_{m \alpha}  \, H^1_{\alpha n}}{(E - E_\alpha)(E_n - E_\alpha)} (E_n - E) \right) \langle n | \phi \rangle \,.
\eeq
Performing an iteration of eq.~(\ref{TDgg4})  
one expresses the last term in the right--hand side of eq.~(\ref{pr1}) as a fourth order term and replace 
$E$ by $E_n$. The final result for the energy independent Tamm--Dancoff potential, 
which differs from the original one in eq.~(\ref{TDpot}) by higher order terms, is then
\beqa
\label{TDen_indep}
(V_{\rm eff})_{mn} &=& g^2 \frac{H^1_{m \alpha} \,  H^1_{\alpha n}}{E_n - E_\alpha}\\
&& {} + g^4 \frac{H^1_{m \alpha}  \, H^1_{\alpha \beta}  \, H^1_{\beta \gamma} \,  
H^1_{\gamma n}}{(E_n - E_\alpha)
(E_n - E_\beta)(E_n - E_\gamma)}
 - g^4 \frac{H^1_{m \alpha}  \, H^1_{\alpha l} \,  
H^1_{l \gamma} 
 \, H^1_{\gamma n}}{(E_n - E_\alpha) (E_l - E_\alpha)(E_n - E_\gamma)}\,.
\nonumber
\eeqa
Even after elimination of the explicit energy dependence, 
eq.~(\ref{TDschroed}) does not correspond
to the  ordinary 
Schr\"odinger equation since the resulting effective Hamiltonian turns out 
to be non--hermitean.
The problem is that the projected nucleon states $| \phi \rangle$ have a 
normalization different from the 
the states $| \Psi \rangle$ we have started from, 
which are assumed to span a complete and orthonormal
set in the whole Fock space:
\beq
\langle \phi_i | \phi_j \rangle =  
\langle \Psi_i | \Psi_j \rangle - \langle \psi_i | \psi_j \rangle=
\delta_{ij} - \langle \phi_i | H_I \lambda 
\left( \frac{1}{E - \lambda H \Lambda} \right)^2 
\lambda H_I \phi_j \rangle ~.
\eeq
Note that the components $\psi_i$ in this equation do, in general, not vanish.
Thus in order to end up with the ordinary Schr\"odinger equation one
has to switch from the set $\{ \phi \}$ of eigenstates of eq.~(\ref{TDschroed}) 
which is assumed to be complete in the $\eta$--subspace\footnote{This is presumably true at least 
in the case of weakly interacting fields.} but is not orthonormal, to another complete 
{\bf and orthonormal} set $\{ \chi \}$. 
This has been achieved in an elegant way by Fukuda et al.
\cite{Fuk54} and Okubo \cite{Okubo54}. 
They introduced an operator $A \equiv \lambda A \eta$ which maps
the states $\phi$ into the states $\psi$:
\beq
\label{defA}
| \psi \rangle = A | \phi \rangle\,,
\eeq
so that the state $\Psi$ is given by
\beq
| \Psi \rangle = ( 1 + A ) | \phi \rangle\,.
\eeq
The operator $A$ is equivalent to the operator which enters eq.~(\ref{5.3})
with the only difference of being energy independent. One can show that it has to
satisfy the equation \cite{Okubo54}:
\begin{equation}
\label{5.10}
\lambda \left( H - \left[ A, \; H \right] - A H A \right) \eta = 0~.
\end{equation}
It is now obvious that the states $| \chi \rangle$ 
\beq
\label{defchi}
| \chi \rangle = (1 + A^\dagger A )^{1/2} |\phi \rangle
\eeq
lie in the $\eta$--subspace and are normalized properly.
The effective Schr\"odinger  equation for $| \chi \rangle$ can now be derived from 
eqs.~(\ref{schroed2}), (\ref{defA}), (\ref{5.10})  and (\ref{defchi}), see ref.~\cite{Okubo54}:
\beq
\label{tempx1}
(H_0 + \tilde{V}_{\rm eff})
| \chi \rangle
= E | \chi \rangle,
\eeq
with the effective potential given by 
\beq
\label{effpot}
\tilde{V}_{\rm eff} =  (1 + A^\dagger A)^{-1/2} (\eta + A^\dagger) 
H (\eta + A )  (1 + A^\dagger A)^{-1/2} - H_0~.
\eeq

A more elegant way to end up with the same effective Schr\"odinger equation has been 
proposed by Okubo by using a unitary transformation of the form 
\begin{equation}
\label{5.9}
U = \left( \begin{array}{cc} \eta (1 + \tilde A^\dagger \tilde A )^{- 1/2} & - 
\tilde A^\dagger ( 1 + \tilde A \tilde A^\dagger )^{- 1/2} \\
\tilde A ( 1 + \tilde A^\dagger \tilde A )^{- 1/2} & 
\lambda (1 + \tilde A \tilde A^\dagger )^{- 1/2} \end{array} \right)~,
\end{equation}
with the operator $\tilde A= \lambda \tilde A \eta$. 
The operator $\tilde A$ has to satisfy 
the decoupling equation (\ref{5.10}) in order for the transformed Hamiltonian 
to be of block--diagonal form
and can (under certain circumstances) be identified with 
the operator $A$ from eq.~(\ref{defA}).   
In \cite{EGM1} we have shown how to solve the decoupling equation (\ref{5.10}) 
and to derive the 
nuclear force according eq.~(\ref{effpot}) using the methods of 
CHPT. In what follows we will 
perform a complete renormalization of the NN potential at NLO  
and demonstrate that the renormalized expressions for the potential in this formalism (with the specific
choice of the unitary operator given in eq.~(\ref{5.9}))
agrees with the one obtained by using the S--matrix methods. 
We will also show how to perform renormalization in the method of unitary transformation
and how to include pion tadpole contributions and recover the same expressions for 
renormalized quantities as found in covariant perturbation theory using the technique of Feynman 
diagrams. It is important to perform a complete renormalization of the effective Hamiltonian and to
take into account all the implicit dependence on the quark (or, equivalently pion) mass.

Before closing this section we would like to stress that the effective 
Hamiltonian derived with the method
of unitary transformation is by no means a unique one, as it has also been the 
case in the S--matrix method. 
The unitary operator (\ref{5.9}) is not the most general one. 
Indeed, it is always possible to perform additional 
unitary transformations in $\eta$ and $\lambda$--subspaces, 
which are not taken into account in the 
definition (\ref{5.9}). The above unitary transformation is in that sense 
the ``minimal'' one. This ambiguity with respect to the effective Hamiltonian
clearly does not mean that the physically observed quantities are defined ambiguously.
In what follows, will not be concerned any more with these subtleties.

\subsection{One--pion exchange contribution to the scattering amplitude}
\def\theequation{\arabic{section}.\arabic{equation}}

\begin{figure}[htb]
\vspace{0.5cm}
\centerline{
\psfig{file=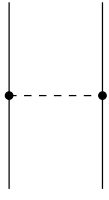,width=2cm}}
\vspace{0.3cm}
\centerline{\parbox{14cm}{
\caption[fig4]{\label{opeLO} 
Leading order one--pion exchange. 
The solid (dashed) lines refer to nucleons 
(pions). The heavy dots are leading order vertices 
from $\mathcal{L}_{\pi N}^{(1)}$.
The diagrams resulting from interchange of the nucleon lines are not shown.
}}}
\vspace{0.7cm}
\end{figure}

\begin{figure}[htb]
\vspace{0.5cm}
\centerline{
\psfig{file=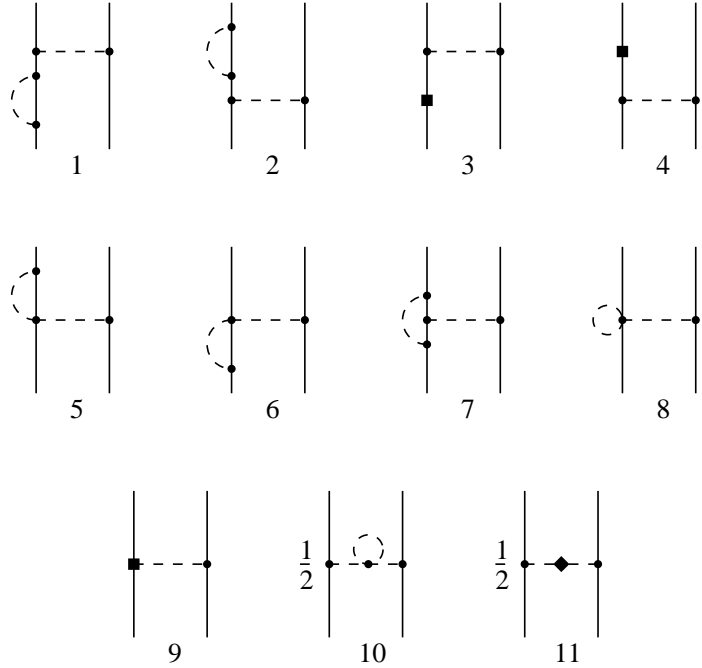,width=11cm}}
\vspace{0.3cm}
\centerline{\parbox{14cm}{
\caption[fig4]{\label{fig1} 
Diagrams contributing to the renormalization of the OPE. 
The heavy dots are leading order vertices 
from $\mathcal{L}_{\pi N}^{(1)}$  and $\mathcal{L}_{\pi \pi}^{(2)}$
while the solid rectangles correspond 
to vertices from $\mathcal{L}_{\pi N}^{(3)}$ 
(in case of diagrams 3 and 4 also from 
$\mathcal{L}_{\pi N}^{(2)}$). The solid diamond represents 
insertions from $\mathcal{L}_{\pi \pi}^{(4)}$.
For remaining notation see fig.~\ref{opeLO}.
}}}
\vspace{0.7cm}
\end{figure}

The contribution from the OPE  to the NN force shown in fig.~\ref{opeLO} appears (in  
Weinberg's power counting) at  leading order (LO) 
in the chiral expansion together with two different contact 
interactions without derivatives.
The OPEP gets renormalized at NLO due to pion loops and counter term insertions. 
As discussed in the previous section, the renormalized OPEP in the 
S--matrix method is defined 
by off--the--energy shell extension of the field--theoretical amplitude.
The corresponding diagrams are shown in 
fig.~\ref{fig1}. The graphs 1--4 lead to the renormalization
of the nucleon lines, while the diagrams 10 and 11 contribute 
to renormalization of the pion line and 
7--9 renormalize the pion--nucleon coupling. The contribution from 
graphs 5 and 6 involves an odd power 
of the loop momentum $l$ to be integrated over and thus vanishes.
The underlying chiral Lagrangians are given by:
\beqa
\label{chirlagr}
\mathcal{L}_{\pi \pi}^{(2)} &=& \frac{F^2}{4} \langle \nabla^\mu U \nabla_\mu 
U^\dagger + \chi_+ \rangle\,,
\nonumber \\
\mathcal{L}_{\pi \pi}^{(4)} &=& \frac{l_3}{16} \langle \chi_+ \rangle^2 \\
&& {} + \frac{l_4}{16} \left\{ 2 \langle \nabla_\mu U \nabla^\mu U^\dagger 
\rangle \langle \chi_+ \rangle
+ 2 \langle \chi^\dagger U \chi^\dagger U + \chi U^\dagger \chi U^\dagger 
\rangle - 4 \langle \chi^\dagger
\chi \rangle - \langle \chi_+ \rangle^2 \right\}
+ \ldots\,,
\nonumber \\
\mathcal{L}_{\pi N}^{(1)} &=&  \bar N_v \left[ i ( v \cdot D ) + \krig{g}_A 
( S \cdot u ) \right] N_v\,,
\nonumber \\
\mathcal{L}_{\pi N}^{(2)} &=&  \bar N_v \bigg[ \frac{1}{2 \krig{m}} (v \cdot D)^2 
- \frac{1}{ 2 \krig{m}}
D \cdot D +
c_1 \langle \chi_+ \rangle \bigg] N_v + \ldots\,,
\nonumber \\
\mathcal{L}_{\pi N}^{(3)} &=&  \bar N_v \bigg[ 
 d_{16} S \cdot u \langle \chi_+ \rangle + i d_{18} S^\mu [ D_\mu , \, \chi_-]
 + 
\tilde d_{28} ( i \langle \chi_+ \rangle v \cdot D + \mbox{h.c.} ) \bigg] N_v 
+ \ldots\,,
\nonumber
\eeqa
where  $F$, $\krig{g}_A$ and $\krig{m}$ refer to the bare pion decay constant, the
bare nucleon axial vector coupling
constant and the bare nucleon mass\footnote{If dimensional regularization is used, 
$F$, $\krig{g}_A$ and $\krig{m}$ coincide with the corresponding constants in the 
chiral limit.} and $c_1$ and $d_i$ are low--energy constants.
The brackets $\langle \, \, \rangle$ 
denote traces in the flavor space. We adopted here the heavy baryon 
formulation for nucleon fields, with $v_\mu$ and $S_\mu= (1/2) 
i \gamma_5 \sigma_{\mu \nu} v^\nu$ denoting the 
four--velocity and spin operator, respectively.
The unitary 2 $\times$ 2 matrix $U$ in the flavor space collects 
the pion fields and is defined in the 
$\sigma$--model gauge, which we will use throughout this work, as:
\beq
\label{sigmodg}
U = \frac{1}{F} \left[ \sqrt{F^2 - \fet{\pi}^2} 
+ i \fet \tau \cdot \fet \pi  \right]\,.
\eeq 
Further,
\beqa
&& u = \sqrt{U}\,, \quad \quad u_\mu = i ( u^\dagger \partial_\mu u 
- u \partial_\mu u^\dagger ) \,, \quad \quad 
D_\mu = \partial_\mu + \Gamma_\mu\,,  \\
&& \Gamma_\mu = \frac{1}{2} [ u^\dagger , \, \partial_\mu u ]\,, \quad \quad
\chi_\pm = u^\dagger \chi u^\dagger \pm u \chi^\dagger u\,, \quad \quad
\chi = 2 B (s + i p)\,, \nonumber
\eeqa
where the quantity $B$ is just a constant.
We do not take into account external vector and axial--vector fields. The physically 
relevant values for scalar and pseudoscalar sources $s$ and $p$ are
\beq
s(x) = \mathcal{M} \,, \quad \quad p(x) = 0\,, 
\eeq
where $\mathcal M$ denotes the quark mass matrix. 
Note that we have only shown  explicitly the terms in eq.~(\ref{chirlagr}) 
we will use in our calculation. 
We also note that $1/\krig{m}$ terms are suppressed in 
Weinberg's approach and would appear one order higher.
Further details on construction and structure of the effective 
Lagrangian can be found in 
\cite{BKMprog,Eck96,Fet98,Fet00,FMMS}.

The OPE at leading order corresponds to the diagram shown in fig.~\ref{opeLO} 
and is given by 
\beq
\label{OPEconv}
\mathcal{A}_{\rm OPE} = \frac{\krig{g}_A{}^2}{F^2}  \, 
(\fet \tau_1 \cdot \fet \tau_2)
\, \frac{1}{q^2-M^2} \, (S_1 \cdot q) (S_2 \cdot q)\,.
\eeq
Here $q$ denotes the momentum transfer of the nucleon, i.e. $q = p ' - p$, where 
$p '$ and $p$ are final and initial nucleon momenta. Further, $M$ is the leading
term in the quark mass expansion of the pion mass.
The more convenient expression in  the rest--frame system of the nucleons\footnote{We are only 
interested in the specific kinematic, in which both nucleons move with 
the same velocity $v_\mu$ and the relative 
momentum is small.}
with $v_\mu = (1, 0, 0, 0)$
and $S^\mu = (0, 1/2 \, \vec \sigma)$, where the $\sigma_i$ are the Pauli spin matrices reads
\beq
\label{amplope}
\mathcal{A}_{\rm OPE}^{\rm NC}= - \frac{\krig{g}_A{}^2}{(2 F)^2} 
\frac{1}{\vec q\, ^2 + M^2} 
(\vec \sigma_1 \cdot \vec q \, )\, (\vec \sigma_2 \cdot \vec q \,) + \mathcal{O}(m^{-2})\,. 
\nonumber
\eeq
We have introduced here the superscript ${\rm NC}$ ($=$ ``noncovariant'')
in order to distinguish between the noncovariant notation of eq.~(\ref{amplope}) and covariant
one  in eq.~(\ref{OPEconv}). The noncovariant notation  is appropriate for calculation of the 
S--matrix  based upon the effective Hamilton operator. Note that the physical values
$g_A$, $F_\pi$ and $M_\pi$ of the axial nucleon coupling, pion decay coupling and pion mass
can be used in the expression for the OPE if one restricts oneself to the LO analysis.

We will now evaluate the NLO corrections to the OPE, which result from the diagrams
shown in fig.~\ref{fig1}. The calculation can easily be performed 
using e.g. the
Feynman rules from \cite{BKMprog,Fet00}. 
The diagrams 1,...,4 renormalize external the nucleon lines and lead to 
\beq
\mathcal{A}_{1,2,3,4}=\left( (Z_N)^2 -1 \right)  \, \mathcal{A}_{\rm OPE}\,. 
\eeq
Here  $Z_N$ is the nucleon Z--factor given by
\beq
\label{nuclzfact}
Z_N = 1 + \Sigma ' (0)\,. 
\eeq
The nucleon self--energy $\Sigma (\tilde \omega)$ ($\tilde \omega = v \cdot k$, 
$k$--nucleon momentum)  is given up to order $Q^2$ ($Q$ refers to a generic small momentum scale)
by 
\beq
\Sigma (\tilde \omega) = - 4 M^2 \left( c_1 + 2 \tilde{d}_{28} 
\tilde \omega \right)
+ 3 i \frac{\krig{g}_A{}^2}{F^2} \int \frac{d^4 l}{(2 \pi)^4}\frac{i}{v \cdot 
(k - l) + i \epsilon}
\frac{i}{l^2 - M^2 + i \epsilon} (S \cdot l) (- S \cdot l)\,.
\eeq
We will now proceed in two ways. First, we apply dimensional regularization and 
the same subtraction as in ref.~\cite{Fet98} and recover the standard results for the amplitude.
Secondly, in order to compare our results with the ones obtained in the method of 
unitary transformation we will go into the rest--frame system with $v_\mu = (1, 0, 0, 0)$ and 
$S^\mu = (0, 1/2 \, \vec \sigma)$, where the $\sigma_i$ are the Pauli spin matrices
and perform  integrations over zeroth
component of the loop momentum $l$ via contour methods. Furthermore, we will 
not restrict ourselves to any specific regularization scheme in that 
case since we are only interested in the general structure of the amplitude. 
The divergent integrals are always to be 
understood as being regularized by some standard methods.
In order to distinguish the results obtained in this way from the ones found using 
dimensional regularization and the subtraction defined in ref.~\cite{Fet98} we will 
denote the latter ones by the superscript $\rm DR$.

For the nucleon $Z$--factor $Z_N$ in eq.~(\ref{nuclzfact}) we find:
\beqa
\label{zn}
Z_N^{\rm DR} &=& 1 - \frac{3 \krig{g}_A{}^2 M^2}{32 \pi^2 F^2} 
\left[ 3 \ln \frac{M}{\lambda} + 1 \right]
- 8 M^2 \tilde d_{28}^r (\lambda)\,, \\
Z_N &=& 1 - 3 \frac{\krig{g}_A{}^2}{8 F^2} J_{13}
- 8 M^2 \tilde d_{28}\,. \nonumber
\eeqa
where we have introduced 
\beq
J_{mn} = \int \frac{d^3 l}{(2 \pi)^3}  \frac{l^{2m}}{\omega_l^n}\,.
\eeq
Clearly, $Z_N^{\rm DR}$ can also be obtained directly from $Z_N$ performing dimensional
regularization in three dimensions of the divergent integral $J_{13}$.

The nucleon mass shift is related to the self energy via
\beq
m_N = \krig{m} + \delta m_N\,, \quad \quad \delta m_N = \Sigma (0)\,.
\eeq
Using dimensional regularization one finds
\beq
\label{nmsDR}
\delta m_N^{\rm DR} = - 4 c_1 M^2 - \frac{3 \krig{g}_A^2 M^3}{32 \pi F^2}\,,
\eeq
while the unregularized expression (after performing integration over the zeroth component $l_0$ of the 
loop momentum) reads:
\beq
\delta m_N =  - 4 c_1 M^2 - \frac{3 \krig{g}_A^2}{8 F^2} J_{12}\,.
\eeq
Notice that the mass of the nucleon does not appear explicitly in the expressions 
for the NN potential at NLO but enters the two--nucleon propagator in the 
Lippmann--Schwinger equation. We will discuss this issue later on. 

For diagrams 7,8,9 in fig.~\ref{fig1} we obtain in dimensional regularization (D.R.)
\beqa
\mathcal{A}_{7,8,9}^{\rm DR}&=&Z_{7,8,9}^{\rm DR} \, \mathcal{A}_{\rm OPE}\,,\\
Z_{7,8,9}^{\rm DR} &=& \frac{2 M^2 L (\lambda) }{F^2} 
+ \frac{M^2}{16 \pi^2 F^2} \left[
(\krig{g}_A{}^2 +2) \ln \frac{M}{\lambda} + \krig{g}_A{}^2 \right] 
+ \frac{8 M^2 d_{16}^r (\lambda)}{\krig{g}_A}-
\frac{4 M^2 d_{18}}{\krig{g}_A} \,, \nonumber
\eeqa
where we used the same definitions for $L (\lambda)$ and $d_{16}^r (\lambda)$ 
as in \cite{Fet98}.
Note that the constant $d_{18}$, which leads to the Goldberger--Treiman discrepancy,
is finite and does not get renormalized. 
Switching to the noncovariant notation and 
performing integration over $l_0$ we find 
the following unregularized expressions for these quantities:
\beqa
\label{z789}
\mathcal{A}_{7,8,9}&=&Z_{7,8,9} \, 
\mathcal{A}_{\rm OPE}^{\rm NC} \,,\\
Z_{7,8,9} &=& -\frac{\krig{g}_A{}^2}{4 F^2} 
\left( 1 - d + \frac{2}{d-1} \right)
J_{13} 
%\int \frac{d^3 l}{(2 \pi)^3} \frac{l^2}{\omega_l^3} 
+ \frac{1}{2 F^2} J_{01}
%\int \frac{d^3 l}{(2 \pi)^3} \frac{1}{\omega_l} 
+ \frac{8 M^2 d_{16}}{\krig{g}_A}-
\frac{4 M^2 d_{18}}{\krig{g}_A} \,, \nonumber
\eeqa
where $d=4$ in the physically relevant case of (3+1) space-time dimensions.

Finally, renormalization of the pion line in graphs 10 and 11 in fig.~\ref{fig1}
leads to
\beq
\label{renpion}
\mathcal{A}_{10,11}=\left( \delta Z_{\pi} + \frac{\delta M^2}{q^2-M^2} \right) 
\, \mathcal{A}_{\rm OPE} \,, 
\eeq
where 
\beqa
\label{renpion2}
Z_\pi &=& 1 + \delta Z_\pi\,, \quad \quad \delta Z_\pi = 
- \frac{\Delta_\pi}{F^2} - \frac{2 M^2}{F^2} l_4 \,, \\
M_\pi^2 &=& M^2 + \delta M_\pi^2\,, \quad \quad \delta M_\pi^2 = 
M^2 \left\{ \frac{\Delta_\pi}{2 F^2}
+ \frac{2 M^2}{F^2} l_3 \right\}
\nonumber \,,
\eeqa
Here, $M_\pi$ is the physical value of the pion mass. 
The last term in the brackets in eq.~(\ref{renpion})
gives the correction to the one--pion exchange due to the shift in the pion mass and should be 
ignored if the physical value of the pion mass is used in the expression (\ref{OPEconv}) for the OPE at LO.
The quantity $\Delta_\pi$ is given in D.R. by 
\beq
\Delta_\pi^{\rm DR}= 2 M^2 \left( L (\lambda) 
+ \frac{1}{16 \pi^2} \ln \frac{M}{\lambda} \right) + \mathcal{O} (d-4)~,
\eeq
and in the previously introduced notation with divergent integrals $J_{ij}$ 
just by $\Delta_\pi= 1/2 \, J_{01}$.
Note that in the calculation of the pion wave function renormalization and mass shift 
one has to take into account the noncovariant pion propagator 
of the form:
\beq
\delta_{ab} \Delta_{\mu \nu} (k) = \int d^4 x \langle 0 | T \partial_\mu \pi_a (x) 
\partial_\nu \pi_b (0) | 0 \rangle = \frac{i k_\mu k_\nu}{k^2 - M^2 + i \eta} 
- i g_{\mu 0} g_{\nu 0}\,,
\eeq
as discussed in \cite{Ger71}. Such noncovariant pieces in the 
pion propagators (as well as noncovariant
pieces in the interaction Hamiltonian arising from time derivatives) 
are usually omitted if one works in dimensional regularization.
This does not lead to wrong results since the missing terms correspond 
to power--law divergences, which 
vanish in D.R.. However, in a general case such noncovariant 
pieces should be taken into account.   
For instance, if the last piece in the above expression is omitted 
and cut--off regularization is used, 
pions become massive in the chiral limit due to the tadpole diagram. 

Summing up the contributions of all graphs in fig.~\ref{fig1} 
we obtain the renormalized OPE as
\beq
\label{OPEren}
\mathcal{A}_{\rm OPE}^{\rm ren} = 
- \frac{1}{4} \left( \frac{g_A}{F_\pi} \right)^2 
\left(1 - \frac{4 M_\pi^2 d_{18}}{g_A} \right) \frac{\fet \tau_1 \cdot \fet \tau_2}
{\vec q\,^2 + M_\pi^2} 
(\vec \sigma \cdot \vec q\,)\, (\vec \sigma \cdot \vec q\,) \,,
\eeq
where $g_A/F_\pi$ is given in D.R. by
\beq
\label{gafpidr}
\frac{g_A}{ F_\pi} = 
\frac{\krig{g}_A}{ F} \left( 1- \frac{g_A^2 M_\pi^2}{4 \pi^2 F_\pi^2} \ln \frac{M_\pi}{\lambda}
- \frac{g_A^2 M_\pi^2}{16 \pi^2 F_\pi^2} - 8 M_\pi^2 \tilde d_{28}^r (\lambda)  
+ \frac{4 M_\pi^2}{g_A} d_{16}^r (\lambda)  - \frac{M_\pi^2}{F_\pi^2} l_4^r (\lambda) \right) 
\eeq
Note that we substituted $\krig{g}_A$ and $F_\pi$  in the brackets 
in the right--hand side of above equations 
by their renormalized values $g_A$ and $F_\pi$, 
which is correct at this order. 
If one is only interested in the real world observables, which correspond to the 
physical value of the pion mass, one can equally well 
rewrite this expression in terms of scale independent LECs $\bar d_{16}$ and $\bar l_4$
as follows:
\beq
\label{gafpidrmpi}
\frac{g_A}{ F_\pi} = 
\frac{\krig{g}_A}{ F} \left( 1 
- \frac{g_A^2 M_\pi^2}{16 \pi^2 F_\pi^2} + \frac{4 M_\pi^2}{g_A}
\bar{d}_{16} - \frac{M_\pi^2}{16 \pi^2 F_\pi^2} \bar{l}_4 \right) \,.
\eeq
The  LECs $\bar d_{16}$ and $\bar l_4$ are defined according to 
 refs.~\cite{Fet98} and \cite{GL84}:
\beqa
\label{defbar}
d_{16}^r (\lambda) &\equiv& d_{16} - \frac{\kappa_{16}}{F^2} L (\lambda) = 
\bar d_{16} - \frac{\kappa_{16}}{(4 \pi F)^2} \ln \frac{\lambda}{M}  \, , \\
l_4^r (\lambda) &\equiv& l_4  - 2 L (\lambda)  = \frac{1}{16 \pi^2} 
\bar l_4 + \frac{1}{8 \pi^2} \ln \frac{M}{\lambda} \, , \nonumber
\eeqa
where $\kappa_{16} = \krig{g}_A (4 - \krig{g_A}{}^2 )/8$.
The renormalized  LEC $\tilde {d}_{28}$ has no finite piece 
(or, in other words, the corresponding 
scale--independent LEC $\tilde {\bar{d}}_{28}$ vanishes):
\beq
\tilde d_{28}^r (\lambda) \equiv \tilde d_{28} 
- \frac{\tilde \kappa_{28}}{F^2} L (\lambda) = 
- \frac{\tilde \kappa_{28}}{(4 \pi F)^2} \ln \frac{\lambda}{M}  \,,
\eeq
with $\tilde \kappa_{28} = -9/16  \krig{g_A}{}^2 $.
Thus the corresponding counter term in the first line of 
eq.~(\ref{zn}) cancels against the logarithmic term. 
We note that it is a convention that the contact terms needed for
renormalization do not have a finite part. In principle, one could
allow for such contributions but that would only lead to a reshuffling
of the finite parts of other LECs. If one insists on working with the
minimal number of independent terms, such a choice has to be taken (as
discussed in more detail in \cite{Fet98}).

Finally, we display the unregularized expression for the ratio $g_A/F_\pi$:
\begin{equation}
\frac{g_A}{ F_\pi} = \frac{\krig{g}_A}{ F} \left( 1 - \frac{g_A^2}{8 F_\pi^2}
\left( 4 - d + \frac{2}{d-1} \right) J_{13} 
- \frac{M_\pi^2}{F_\pi^2} l_4 - 8 M_\pi^2 
\tilde d_{28} + \frac{4 M_\pi^2}{g_A} d_{16} \right)\,.
\end{equation}

\subsection{One--pion exchange from the method of unitary transformations}
\def\theequation{\arabic{section}.\arabic{equation}}

Let us now calculate the renormalized OPEP using the method of unitary transformation, following
the lines of refs.~\cite{EGM1,EGM2}. The essential difference compared to our previous work 
\cite{EGM1} is that we now have to take into account pion tadpole contributions. The corresponding 
diagrams lead to renormalization of various parameters in the effective Lagrangian and were of 
no importance in the previous calculations \cite{EGM2}, where physical values for renormalized coupling
constants have been used. The pion tadpole contributions are, however, 
important for the present analysis,
since they lead to an additional pion mass dependence of various renormalized quantities.
In the following we will show how to deal with the pion tadpoles and how to perform complete
and consistent renormalization in the method of unitary transformation. We would also like to stress that
renormalization of the effective Hamilton operator within the method of unitary transformation has been  
considered recently in a different context, see refs.~\cite{Kr99,Kr00}. 

Let us begin with the effective Lagrangian $\mathcal{L}_{\pi \pi}$ for pions given in the 
first two lines of eq.~(\ref{chirlagr}). 
The relevant terms in the $\sigma$--model gauge, eq.~(\ref{sigmodg}), read:
\beqa
\label{lagrpion}
\mathcal{L}_{\pi \pi}^{(2)} &=& \frac{1}{2} \left( \partial_\mu \fet \pi \cdot \partial^\mu \fet \pi 
- M^2 \fet \pi^2 \right) + \frac{1}{2 F^2} \left( \partial_\mu \fet \pi \cdot \fet \pi
\right)^2 - \frac{1}{8 F^2} M^2 \fet \pi^4 + \ldots \\
\mathcal{L}_{\pi \pi}^{(4)} &=& - \frac{l_3 M^4}{F^2} \fet \pi^2 + \frac{l_4 M^2}{F^2} \partial_\mu 
\fet \pi \cdot \partial^\mu \fet \pi - \frac{l_4 M^4}{F^2} \fet \pi^2 + \ldots\,.
\nonumber
\eeqa
We now introduce the renormalized pion field and mass:
\beqa
\label{pionrena}
\fet \pi_r &=& Z_\pi^{-1/2} \, \fet \pi\,, \quad \quad Z_\pi = 1 + \delta Z_\pi\,, \\
M_\pi^2 &=& M^2 + \delta M_\pi^2\,, \nonumber
\eeqa
where $\delta Z_\pi, \, \,(\delta M_\pi^2)/M_\pi^2 \sim \mathcal{O} (Q^2/\Lambda_\chi^2)$.
We express eq.~(\ref{lagrpion}) in terms of the renormalized pion field and mass by simply
replacing $\fet \pi \rightarrow \fet \pi_r$, $M^2 \rightarrow M_\pi^2$ and 
$\mathcal{L}_{\pi\pi}^{(4)} \rightarrow  \mathcal{L}_{\pi\pi}^{(4)} + \delta \left(\mathcal{L}_{\pi\pi}^{(4)}
\right)$ where
\beq
\delta \left(\mathcal{L}_{\pi\pi}^{(4)} \right) = \frac{1}{2} \delta Z_\pi 
\partial_\mu \fet \pi_r \cdot \partial^\mu \fet \pi_r  - \frac{1}{2} M_\pi^2 \left( \delta Z_\pi - 
\frac{\delta M_\pi^2}{M_\pi^2} \right) \fet \pi_r^2 + \ldots\,.
\eeq
Here the ellipses refer to terms with more pion fields.
In what follows we will always work with renormalized pion fields and therefore omit 
the subscript $r$.

The Hamilton density corresponding to the Lagrangian introduced above can be derived using the 
canonical formalism and following the lines of 
\cite{Ger71}. Let us first express the effective Lagrangian in a more compact form:
\beq
\mathcal{L}_{\pi\pi} = \frac{1}{2} \partial_\mu \pi_a G_{ab} (\fet \pi) \partial^\mu \pi_b - \frac{M_\pi^2}{2}
\fet \pi^2 B(\fet \pi)\,,
\eeq
where 
\beqa
G_{ab} (\fet \pi) &=& \delta_{ab} \left( 1 + \delta Z_\pi + \frac{2 l_4 M_\pi^2}{F^2} \right)+ 
\frac{1}{F^2} \pi_a \pi_b  + \ldots \, = \delta_{ab} + \tilde G_{ab} (\fet \pi), \\
B (\fet \pi) &=& 1 + \delta Z_\pi - \frac{\delta M_\pi^2}{M_\pi^2} + \frac{2 l_3 M_\pi^2}{F^2} 
+ \frac{2 l_4 M_\pi^2}{F^2} + \frac{1}{4 F^2} \fet \pi^2 + \ldots = 1 + \tilde B (\fet \pi )\,.
\nonumber
\eeqa
Here the subscripts $a$, $b$ correspond to isospin indices. For the momentum $\Pi_a$ conjugate to the 
field $\pi_a$ we get:
\beq
\Pi_a = \frac{\delta \mathcal{L_{\pi\pi}}}{\delta \partial_0 \pi_a} = G_{ab} (\fet \pi) \partial_0 \pi_b\,.
\eeq
The Hamiltonian $\mathcal{H_{\pi\pi}}$ is given by
\beq
\mathcal{H}_{\pi \pi}=\Pi_a G^{-1}_{ab}(\fet \pi) \Pi_b - \mathcal{L_{\pi\pi}} = \frac{1}{2} 
\Pi_a G^{-1}_{ab}(\fet \pi) \Pi_b - \frac{1}{2} \partial_i \pi_a G_{ab} (\fet \pi) \partial^i \pi_b
+ \frac{M_\pi^2}{2} \fet \pi^2 B(\fet \pi)\,.
\eeq
We now divide the Hamiltonian into its free and interaction part as follows:
\beqa
\label{prtr}
\mathcal{H}_{\pi \pi}&=&\mathcal{H}_{\pi \pi}^0 + \mathcal{H}_{\pi \pi}^{\rm I} \,, \nn
\mathcal{H}_{\pi \pi}^0&=&\frac{1}{2} \Pi_a \Pi_a - \frac{1}{2} \partial^i \pi_a \partial_i \pi_a + 
\frac{M_\pi^2}{2} \fet \pi^2\,, \\
\mathcal{H}_{\pi \pi}^{\rm I}&=& - \frac{1}{2} \partial_\mu \pi_a \tilde G_{ab} (\fet \pi) \partial^\mu \pi_b
- \frac{1}{2} \partial_0 \pi_a \tilde G_{ab}^2 (\fet \pi ) \partial_0 \pi_b
+ \frac{1}{2} M_\pi^2 \fet \pi^2 \tilde B (\fet \pi )\,. \nonumber
\eeqa
Since the field operators $\pi_a$, $\partial_i \pi_a$ and $\Pi_a$ transform simply into the 
interaction picture (IP), $\pi_a \rightarrow \pi_a^{\rm IP}$, 
$\partial_i \pi_a \rightarrow ( \partial_i \pi_a)^{\rm IP}$ and $\Pi_a \rightarrow ( \partial_0 \pi_a )^{\rm IP}$,
we can easily express the interaction Hamiltonian $\mathcal{H}_{\pi \pi}^{\rm I}$ in the IP representation 
as:\footnote{In the last line of eq.~(\ref{prtr}) one has first to replace $\partial_0 \pi_a$ by $G_{ab}^{-1}
    (\fet \pi ) \Pi_b$.} 
\beqa
\label{hamfin}
\mathcal{H}_{\pi\pi}^{\rm I} &=& - \frac{1}{2} \partial_\mu \pi_a \tilde G_{ab} (\fet \pi ) \partial^\mu \pi_b
+ \frac{1}{2} M_\pi^2 \fet \pi^2 \tilde B (\fet \pi) + \frac{1}{2} \partial_0 \pi_a \left\{
\tilde G^2 (\fet \pi)  (1 + \tilde G (\fet \pi) )^{-1} \right\}_{ab} \partial_0 \pi_b \\
&=&- \frac{1}{2} \partial_\mu \fet \pi \cdot \partial^\mu \fet \pi \left( \delta Z_\pi + 
\frac{2 l_4 M_\pi^2}{F^2} \right) + \frac{1}{2} M_\pi^2 \fet \pi^2 \left( \delta Z_\pi - 
\frac{\delta M_\pi^2}{M_\pi^2} + \frac{2 l_3 M_\pi^2}{F^2} + \frac{2 l_4 M_\pi^2}{F^2} \right)\nn
&& {} - \frac{1}{2 F^2} \left( \fet \pi \cdot \partial_\mu \fet \pi \right)
\left( \fet \pi \cdot \partial^\mu \fet \pi \right) + \frac{1}{8 F^2} M_\pi^2 \fet \pi^4 + \ldots \,,
\nonumber 
\eeqa
where we have omitted the superscripts $\rm IP$. We now write $\pi_a (x)$ in terms of creation and annihilation 
operators $a^\dagger_a (\vec k)$, $a_a (\vec k)$ as
\beq
\label{secquant}
\pi_a (x) = \int \, \frac{d^3 k}{(2 \pi)^{3/2}} \, \frac{1}{\sqrt{2 \omega_k}} \left[ e^{-i k \cdot x} a_a (\vec k)
+ e^{i k \cdot x} a_a^\dagger (\vec k) \right]\,,
\eeq
where $\omega_k = \sqrt{\vec k \, ^2 + M_\pi^2}$ and the operators $a_a(\vec k)$ and $a_a^\dagger (\vec k)$
satisfy the commutation relations
\beq
\left[ a_a (\vec k ), \, \,  a_b (\vec k\, ' ) \right] = \left[ a_a^\dagger 
(\vec k ), \, \,  a_b^\dagger (\vec k\, ' ) \right] = 0, \quad \quad 
\left[ a_a (\vec k ), \, \,  a_b^\dagger (\vec k\, ' ) \right] = \delta_{ab} \, \delta (\vec k - \vec k \, ')\,.
\eeq 
\begin{figure}[htb]
\vspace{1.2cm}
\centerline{
\psfig{file=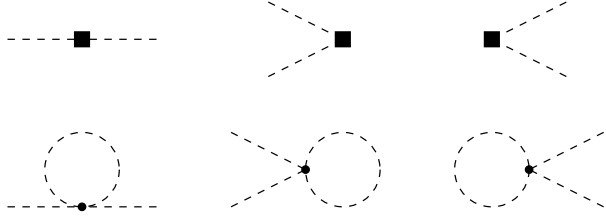,width=9cm}}
\vspace{0.3cm}
\centerline{\parbox{14cm}{
\caption[fig4]{\label{tadpol}
Contributions to the one--pion Hamiltonian.
The heavy dots are leading order vertices from $\mathcal{L}_{\pi\pi}^{(2)}$ 
while the solid rectangles correspond 
to vertices from $\mathcal{L}_{\pi\pi}^{(4)} + \delta \left( \mathcal{L}_{\pi\pi}^{(4)}
\right)$.
}}}
\vspace{0.7cm}
\end{figure}
Substituting eq.~(\ref{secquant}) into eq.~(\ref{hamfin}) and
performing normal ordering,
 we end up with  terms of the form $a^\dagger a$, $a^\dagger a^\dagger$, $a a$,
$a^\dagger a^\dagger a a$, $\ldots$. 
In fig.~\ref{tadpol} we show symbolically the contributions to the Hamiltonian of the form 
$a^\dagger a$, $a^\dagger a^\dagger$ and $a a$. Note that the closed loops result
from contraction of the operators $a$ and $a^\dagger$
in the terms in the 
last line of eq.~(\ref{hamfin}) when bringing these into normal ordered form.
For example, for the term of the form $a^\dagger a$ we get the following
expression:
\beqa
\label{pionreno}
H_{\pi\pi}^{\rm I} &=&  \sum_{a}\,\int d^3 k \,\frac{1}{\omega_k}\, a_a^\dagger (\vec k) a_a (\vec k)
\bigg\{\frac{1}{2} k^2 \left( - \delta Z_\pi- \frac{2 l_4 M_\pi^2}{F^2} - \frac{1}{2 F^2} J_{01} \right)
\\
&& {} \quad \quad \quad \quad \quad \quad \quad \quad \quad \quad \quad \quad
+ \frac{1}{2} M_\pi^2 \left(  \delta Z_\pi - \frac{\delta M_\pi^2}{M_\pi^2} + \frac{ 2 l_3 M_\pi^2}{F^2}
+\frac{2 l_4 M_\pi^2}{F^2} + \frac{3}{4 F^2} J_{01} \right) \bigg\} + \ldots\,.
\nonumber
\eeqa
We now require that there are no terms of the form $a^\dagger a$, $a^\dagger a^\dagger$ and $a a$
in the interaction Hamiltonian at the order considered in the chiral expansion (since we already 
work with the renormalized pion fields). Then we obtain from eqs.~(\ref{pionrena}),(\ref{pionreno}):
\beqa
Z_\pi &=& 1 - \frac{2 l_4 M_\pi^2}{F^2} - \frac{1}{2 F^2} J_{01}\,, \\
M_\pi^2 &=& M^2 \left( 1 + \frac{2 l_3 M_\pi^2}{F^2} + \frac{1}{4 F^2} J_{01} \right)\,.
\nonumber
\eeqa
These expressions coincide precisely with the ones found using the S--matrix method, 
eq.~(\ref{renpion2}).

Let us now include nucleons. The relevant terms in the effective Lagrangian $\mathcal{L}_{\pi N}$ from
eq.~(\ref{chirlagr}) are
\beqa
\label{lagrpin}
\mathcal{L}_{\pi N}^{(1)} &=& N^\dagger \left(i \partial_0  -   \frac{\krig{g}_A}{2 F} \fet \tau \vec \sigma
\cdot \vec \nabla \fet \pi - \frac{\krig{g}_A}{4 F^3} (\fet \pi \vec \sigma \cdot \vec \nabla \fet \pi )
\fet \tau \cdot \fet \pi ) \right) N + \ldots\,,
%- \frac{1}{4 F^2} \fet \tau \cdot (\fet \pi \dot{\fet \pi} ) \right) + \ldots\,,
\nn
\mathcal{L}_{\pi N}^{(2)} &=& N^\dagger \left( \delta m_N + \frac{\vec \nabla^2}{2 m_N} + 4 c_1 M_\pi^2 
\right) N + \ldots \,, \\
\mathcal{L}_{\pi N}^{(3)} &=& N^\dagger \left( \left( -\frac{2 M_\pi^2 d_{16}}{F} + 
\frac{M_\pi^2 d_{18}}{F} - \frac{\krig{g}_A}{4 F} \delta Z_\pi  \right)
\fet \tau \vec \sigma \cdot \vec \nabla \fet \pi + 8 i \tilde d_{28} M_\pi^2 \partial_0 \right) N + \ldots \,.
\nonumber
\eeqa
Here several comments are in order. First, the superscript $(i)$ in the above Lagrangians refers, as in 
eq.~(\ref{chirlagr}), just to the number of derivatives and/or pion mass insertion. 
Secondly, we prefer to use the physical and not the 
bare mass of the nucleon in the heavy baryon expansion, i.e. we parameterize the four--momentum of the nucleon
$p_\mu$ as $p_\mu = m_N v_\mu + l_\mu$, where $l_\mu$ is a small residual momentum $v \cdot l \ll m_N$.
Expressing the Lagrangian in terms of the physical instead of the bare nucleon mass 
is, of course, just a matter of convenience and all results remain unchanged.
The physical and bare nucleon masses are related by
\beq
m_N = \krig{m} + \delta m_N\,, \quad \quad \delta m_N = \delta m_N^{(2)} + \delta m_N^{(3)} + \ldots\,,
 \quad \quad  \delta m_N^{(i)}  \sim \mathcal{O}\left(Q^i \right)\,.
\eeq
Note that $\delta m_N^{(2)} = - 4 c_1 M_\pi^2$ as can be seen from the second line of 
eq.~(\ref{lagrpin}).\footnote{Pion loops start to contribute at next higher order in the chiral
expansion.}
As in the previous section 
we restrict ourselves to the rest--frame system of the nucleons with $v_\mu = (1, 0, 0, 0)$. 
Notice that the 
term $N^\dagger \delta m_N N$ enters the Lagrangian $\mathcal{L}_{\pi N}^{(2)}$ 
as a consequence of using the physical nucleon mass instead of the bare one. 
Further, we have not included the term $-1/(4 F^2 ) N^\dagger \fet \tau
\cdot (\fet \pi \times \dot{\fet \pi} ) N$  
in the effective Lagrangian $\mathcal{L}_{\pi N}^{(1)}$, which results from the covariant derivative 
of the nucleon field in the third line of eq.~(\ref{chirlagr}). As already explained above, the 
corresponding contributions to the OPE vanish since an odd power 
of the loop momentum $l$  to be integrated over enters the explicit expressions.
Finally, it should be kept 
in mind that the effective Lagrangian (\ref{lagrpin}) is expressed in terms of renormalized pion fields,
which leads to the additional contribution 
with an insertion of $\delta Z_\pi$ in $\mathcal{L}_{\pi N}^{(3)}$.
The Hamilton density corresponding to eq.~(\ref{lagrpin}) is given by
\beqa
\label{hampin}
\mathcal{H}_2^0 &=& -\tilde N^\dagger \frac{\vec \nabla^2}{2 m_N} \tilde N\,, \nn
\mathcal{H}_0 &=& \tilde N^\dagger \left(
\frac{\krig{g}_A}{2 F} \fet \tau \vec \sigma \cdot \vec \nabla \fet \pi+ 
\frac{\krig{g}_A}{4 F^3} (\fet \pi \vec \sigma \cdot \vec \nabla \fet \pi )
\fet \tau \cdot \fet \pi   \right)\tilde N+ \ldots \,, \\
\mathcal{H}_2 &=& \tilde N^\dagger \left(- \delta m_N^{(3)}+ \frac{2 M_\pi^2 d_{16}}{F} - 
\frac{M_\pi^2 d_{18}}{F}+ \frac{\krig{g}_A}{4 F} \delta Z_\pi  \right) 
\fet \tau \vec \sigma \cdot \vec \nabla \fet \pi
\tilde N+ \ldots \,.
\nonumber
\eeqa
Here we switched to the field $\tilde N$ defined as 
\beq
\label{norm1}
\tilde N = (Z_N^N)^{-1/2} N = \left( 1 - \frac{\delta Z_N^N}{2} \right) = (1 + 4 M_\pi^2 \tilde d_{28} ) N\,,
\eeq
in order to get rid of the time derivative in the third line of eq.~(\ref{lagrpin}).
Further we have used a classification of the vertices in effective Lagrangian different from ref.\cite{EGM1}.
To be more precise, the dimension $\Delta$ of the interaction $\mathcal{H}_\Delta$ is defined as:
\beq
\Delta = d + \frac{1}{2} n  + l - 2\,.
\eeq
Here $d$, $n$ and $l$ is the number of derivatives or insertions of $M_\pi$, nucleon  
field operators and insertions of the factors $1/m_N$, respectively. 
According to the power counting suggested by Weinberg \cite{wein}
we treat the nucleon mass as a much larger scale compared to $\Lambda_\chi$ ($m_N \sim \Lambda_\chi^2/Q$).
As a consequence, relativistic corrections enter at higher orders compared to the corresponding chiral
corrections. Notice further that the nucleon field operators in eq.~(\ref{hampin}) are always taken in 
the normal ordering. Allowing contractions of nucleons field operators would only lead to shifts
in the values
of various coupling constants in the effective Lagrangian (Hamiltonian) which are independent of 
$M_\pi$ and thus of no importance for our considerations.\footnote{Remember that 
one does not start from a normal ordering 
for 
pion field operators is required. Pion tadpole diagrams lead to $M_\pi$--dependent renormalization
of various constants in the effective Lagrangian and thus must be considered explicitly.}

We will now follow the lines of \cite{EGM1} and derive the effective Hamiltonian, which acts on the purely 
nucleonic subspace of the whole Fock space, using the method of unitary transformation.  The idea 
of this method is described in section \ref{sec:constrpot}.
For a detailed discussion on the way of solving the nonlinear decoupling equation (\ref{5.10})
and calculating the effective Hamiltonian according to eqs.~(\ref{tempx1}),(\ref{effpot})
the reader is addressed to ref.~\cite{EGM1}. 

Let us begin with the one--nucleon effective Hamiltonian $\tilde H_{\rm 1N}$:
\beqa
\tilde H_{\rm 1N} &=& \tilde H_{\rm 1N}^{(0)} + \tilde H_{\rm 1N}^{(1)}  
+ \ldots \,, \\
\tilde H_{\rm 1N}^{(0)} &=& \eta_1 \left( H_2  + A^\dagger_0 \lambda^1 H_0 
+ H_0 \lambda^1 A_0 + A^\dagger_0 
\lambda^1 \omega A_0  \right) \eta_1 \nn
                        &=& \eta_1 \left( H_2   - H_0 \frac{\lambda^1}{\omega} H_0 \right) \eta_1 \,, \nn
&\ldots&\,,
\nonumber
\eeqa
where the subscript $i$ (superscript $j$) of the projector $\eta_i$ ($\lambda^j$) denotes the number of 
nucleons (pions) in the corresponding state and $\omega$ is the pionic free energy. 
Further, $H_i$ denote Hamilton operators corresponding to the Hamilton density $\mathcal{H}_i$
in eq.~(\ref{hampin}).
The operator
$\lambda^1 A_0 \eta_1$ is given by \cite{EGM1}:
\beq
\lambda^1 A_0 \eta_1 = - \frac{\lambda^1}{\omega} H_0 \eta_1\,.
\eeq
The superscript $\nu$ of $\tilde H_{\rm 1N}^{(\nu)}$ refers to the order in the chiral expansion and
is defined as:\footnote{We use here the definition of the counting index $\nu$ different from the one 
introduced in ref.~\cite{EGM1}. The presently used notation is more transparent when  
operators with different number of nucleons, i.e. 1N, 2N, 3N, $\ldots$ forces, are considered.
All results of ref.~\cite{EGM1} remain however unchanged.}
\beq
\nu = -2 + 2 E_n + 2 ( L - C) + \sum_i V_i \Delta_i\,,
\eeq
where $E_n$, $L$, and $C$ denote the number of nucleons, closed loops and separately
connected pieces, in order. Furthermore, $V_i$ is the number of vertices of type $i$.  
Performing straightforward calculations we end up with the following result:
\beq
\label{ham1n}
h_{\rm 1N}^{(0)} = \frac{\vec p\, ^2}{2 m_N} - \delta m_N^{(3)}- \frac{3 \krig{g}_A^2}{8 F^2} J_{12} \,,
\eeq
where we switched to a more convenient nonrelativistic notation, i.e. we omite creation and annihilation 
operators as well as summation over the appropriate quantum numbers. To be more precise, the relation 
between $H_{\rm 1N}$ and $h_{\rm 1N}$ is given by 
\beq
H_{\rm 1N} = \sum_\alpha \int \, d^3 p \, 
n^\dagger_\alpha (\vec p\,)  \, n_\alpha (\vec p\,) \, h_{\rm 1N}
(\vec p\,)~,
\eeq  
where $\alpha$ refers to discrete 
quantum numbers (spin and isospin) and $n^\dagger_\alpha (\vec p\,)$  ($n_\alpha (\vec p\,)$) is a 
creation (annihilation) operator for the nucleon field. It is now easy to read off the expression
for the nucleon mass shift $\delta m_N$ up to the order considered here:
\beq
\label{nms}
\delta m_N= - 4 c_1 M_\pi^2 - \frac{3 \krig{g}_A^2}{8 F^2} J_{12}\,,
\eeq
which agrees precisely with the result obtained in the previous section using the covariant
perturbation theory. One sees from eq.~(\ref{ham1n}) that the leading one--nucleon 
Hamiltonian is given, in accordance with our expectation, just
by the nucleon kinetic energy $h_{\rm 1N}^{(0)} = \vec p\, ^2 \,/(2 m_N)$. 
Notice that since we are interested in the NN interaction at order $\nu = 2$, 
we should in principle also consider  $h_{\rm 1N}^{(2)}$ and thus include higher order 
nucleon mass shifts. We will, however, see explicitly in what follows that this is not
necessary. Moreover, we will show that even the leading shift in nucleon mass,
$\delta m_N^{(2)} = - 4 c_1 M_\pi^2$, contributes at NNLO and thus can be neglected.

The nucleon $Z$--factor $Z_N$ needs, in principle, not be calculated separately in the method of 
unitary transformation, since it already enters eq.~(\ref{effpot}). More precisely,
the part of the $Z$--factor $Z_N^\pi$, which corresponds to dressing of the bare nucleon by virtual pions, 
is related to
the operator $\eta_1 (1 + A^\dagger A)^{- 1/2} \eta_1$. Indeed, the $Z$--factor is given by
\beq
\label{tempxa}
| \chi_1 \rangle = (Z_N^\pi )^{-1/2} | \phi_1 \rangle\,,
\eeq
where $| \phi_1 \rangle$ is the original (bare) one--nucleon state and 
$| \chi_1 \rangle$ is the properly normalized one--nucleon state (i.e.~which satisfies
the orthonormality condition as described in section \ref{sec:constrpot}). It follows from 
eqs.~(\ref{tempxa}),(\ref{defchi}) that:
\beqa
(Z_N^\pi)^{-1/2} &\equiv& \langle \phi_1 | \chi_1 \rangle 
= \langle \phi | \eta_1 ( 1 + A^\dagger A)^{1/2} \eta_1 | \phi \rangle \\
&=& \langle \phi | \eta_1 \left( 1 + \frac{1}{2} A^\dagger_0 \lambda^1 A_0  + \ldots \right) 
\eta_1 | \phi \rangle 
=  1 + \frac{3 \krig{g}_A^2}{16 F^2} J_{13} + \ldots \,.
\nonumber
\eeqa 
The complete nucleon $Z$--factor $Z_N$, which is related to the specific choice of nucleon field made in 
the Lagrangian (\ref{lagrpin}), is given by (at the order considered) 
\beq
Z_N = Z_N^N Z_N^\pi = 1 - 8 M_\pi^2 \tilde d_{28} - \frac{3 \krig{g}_A^2}{8 F^2} J_{13}\,.
\eeq 
This agrees with $Z_N$ from eq.~(\ref{zn}). We should, however, stress that the
$Z$--factor is not an observable quantity and depends on the choice of fields. For example, there 
would be no contribution proportional to $\tilde d_{28}$ in $Z_N$ if we had decided to work with the fields 
$\tilde N$ instead of $N$.
For more discussion on the role of the nucleon $Z$--factor in the old--fashioned perturbation theory 
and Hamiltonian formalism the reader is referred to  \cite{Okubo54,Fuk54,GML54,Chew54,Wick55} (for
corresponding discussions in the framework of baryon CHPT, see e.g. \cite{EM,FMSwf}).

Let us now consider the 2N effective Hamiltonian at order $\nu=2$. 
\begin{figure}[htb]
\vspace{0.5cm}
\centerline{
\psfig{file=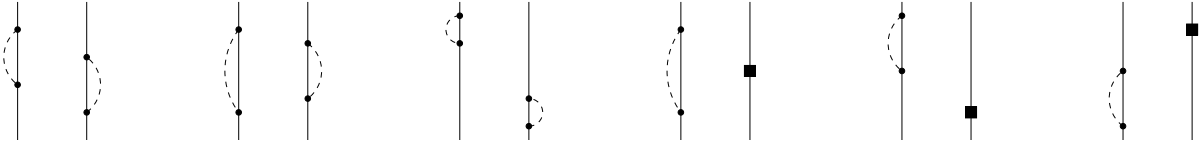,width=16.5cm}}
\vspace{0.3cm}
\centerline{\parbox{14cm}{
\caption[fig4]{\label{discon}
Contributions to the two--nucleon Hamiltonian at order $\nu=2$: disconnected graphs.
Heavy dots refer to the leading order $\pi NN$ vertex of dimension $\Delta=0$ 
while the solid rectangles correspond 
to insertions of the vertex with $\Delta=2$. For remaining notation see fig.~\ref{opeLO}.
}}}
\vspace{0.7cm}
\end{figure}
The explicit operators at NLO are given in ref.~\cite{EGM1}. 
For our purposes we do not need all terms but only the following ones:
\beqa
\label{tempp3}
h_{\rm 2N}^{(2)}&=& \eta_2 \bigg(
- H_0 \frac{\lambda^1}{\omega} H_0 \frac{\lambda^2}{\omega_1 +
  \omega_2} H_0 
\frac{\lambda^1}{\omega} H_0 + \frac{1}{2} H_0
\frac{\lambda^1}{(\omega )^2} H_0 \eta_2 
H_0 \frac{\lambda^1}{\omega} H_0 + \frac{1}{2} H_0 \frac{\lambda^1}
{\omega} H_0 \eta_2 
H_0 \frac{\lambda^1}{(\omega )^2} H_0 
\nn
&&{} \quad \quad + H_0  \frac{\lambda^1}{\omega} 
H_2 \frac{\lambda^1}{\omega} H_0  
- \frac{1}{2}
H_0 \frac{\lambda^1}{( \omega )^2} H_0 \eta_2 H_2  - \frac{1}{2} H_2 \eta_2 H_0  
\frac{\lambda^1}{( \omega )^2} H_0 \bigg) \eta_2~.
\eeqa
In fig.~\ref{discon} we show disconnected diagrams, which contribute to the 2N effective Hamiltonian
at order $\nu=2$, which do not include the nucleon 
self--energy.\footnote{Notice that the third diagram in 
fig.~\ref{discon} is related to the last two operators in the first line of eq.~(\ref{tempp3}) and thus 
does not correspond to the iteration of the nucleon self--energy contribution. The same holds true for last 
two graphs in this figure.}
Denoting by ${\cal M}$ the common spin--isospin structure as well as the loop integrals 
and using the  first line of eq.~(\ref{tempp3})
we see that the contribution of the first three diagrams in fig.~\ref{discon} vanishes:
\beq
{\cal M} \, \left(- \frac{2}{\omega_1^2 \omega_2^2 (\omega_1 + \omega_2)} - 
\frac{1}{\omega_1^3 \omega_2 (\omega_1 + \omega_2)}- 
\frac{1}{\omega_1 \omega_2^3 (\omega_1 + \omega_2)} + \frac{1}{\omega_1^3 \omega_2^2}
+ \frac{1}{\omega_1^2 \omega_2^3} \right)=0
\eeq
A similar cancellation is also  observed for the remaining graphs in fig.~\ref{discon}
using the second line of eq.~(\ref{tempp3}). Thus no disconnected diagrams (which are different
from the nucleon self energy graphs) contribute to the NN potential at NLO.

\begin{figure}[htb]
\vspace{0.5cm}
\centerline{
\psfig{file=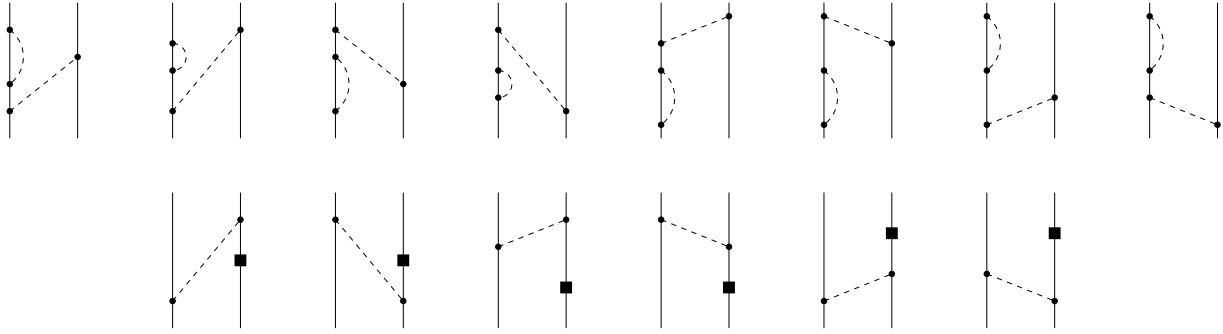,width=17.5cm}}
\vspace{0.3cm}
\centerline{\parbox{14cm}{
\caption[fig4]{\label{ope1}
Contributions to the OPE at order $\nu=2$.
For notation see fig.~\ref{discon}.
}}}
\vspace{0.7cm}
\end{figure}

Consider now the diagrams shown in fig.\ref{ope1}, which are of the same topology as Feynman  
graphs 1--4 in fig.~\ref{fig1}.
The corresponding operators are given in eq.~(\ref{tempp3}).
Note that the contribution of the diagrams in the lower row in fig.~\ref{ope1} vanishes similarly
to the previously considered case of disconnected graphs. Performing explicit calculations one 
finds the same result as in the covariant approach, namely:\footnote{In 
order to be consistent with the results of the previous section
we give here and in what follows 
explicit expressions for the effective Hamiltonian based upon the nucleon field  
$N$, which is related to $\tilde N$ via eq.~(\ref{norm1}). As a consequence, we have to 
take into account the additional normalization factor $(Z_N^N)^{-1/2}$.}
\beq
h_{\rm 2N}^{(2)} = Z_N \tilde \mathcal{A}_{\rm OPE}^{\rm NC}\,,
\eeq
where $Z_N$ is given in eq.~(\ref{zn}). The OPE amplitude 
$\tilde \mathcal{A}_{\rm OPE}^{\rm NC}$ is the same as $\mathcal{A}_{\rm OPE}^{\rm NC}$
in eq.~(\ref{amplope}) with the only difference that the physical pion mass $M_\pi$
is used instead of $M$ (since we are working now with renormalized pion fields). 
We will not consider the topologies related to graphs 5 and 6 in fig.~\ref{fig1}, since the 
corresponding contributions vanish as explained above. 
\begin{figure}[htb]
\vspace{0.5cm}
\centerline{
\psfig{file=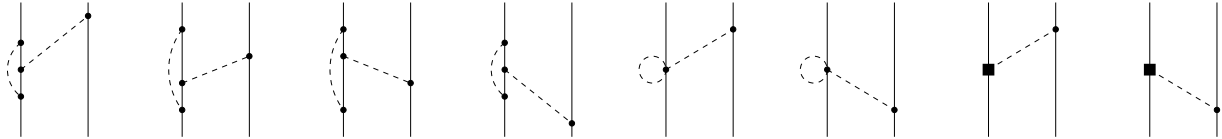,width=17.5cm}}
\vspace{0.3cm}
\centerline{\parbox{14cm}{
\caption[fig4]{\label{ope2}
Remaining contributions to the OPE at order $\nu=2$.
For notation see fig.~\ref{discon}.
}}}
\vspace{0.7cm}
\end{figure}
The remaining contributions to the OPE at NLO in the method of unitary transformation are 
depicted in fig.~\ref{ope2}.
The first four diagrams show all possible time orderings of the covariant graph
7 of fig.~\ref{fig1} and correspond to the first operator in eq.~(\ref{tempp3}). Explicit evaluation
leads to the same expression as found in the covariant approach, see eq.~(\ref{z789}):
\beq
h_{\rm 2N}^{(2)} = Z_7 \tilde \mathcal{A}_{\rm OPE}^{\rm NC}\,,
\quad \quad  \quad  Z_7=
-\frac{\krig{g}_A{}^2}{4 F^2} 
\left( 1 - d + \frac{2}{d-1} \right)
J_{13} \,.
\eeq
The two pion tadpole diagrams in fig.~\ref{ope2} result from contracting two pion field operators
when expressing the term $-\tilde N^\dagger 
\frac{\krig{g}_A}{4 F^3} (\fet \pi \vec \sigma \cdot \vec \nabla \fet \pi )
\fet \tau \cdot \fet \pi \tilde N$ from eq.~(\ref{hampin}) in normal ordered form, as it also 
happened in case of the pion self--energy discussed above. A straightforward but somewhat tedious calculation
leads again to the same result as in the covariant approach:
\beq
h_{\rm 2N}^{(2)} = Z_8 \tilde \mathcal{A}_{\rm OPE}^{\rm NC}\,,
\quad \quad  \quad  Z_8=\frac{1}{2 F^2} J_{01}
\,.
\eeq
Consider now the last two diagrams in fig.~\ref{ope2}. Since we have chosen to work with 
renormalized pion field operators in the method of unitary transformation, we have no diagrams 
corresponding to graphs 10 and 11 in fig.~\ref{fig1}. Contributions of these two graphs are now
taken into account by using the physical pion mass $M_\pi$ instead of $M$ 
as well as by an additional interaction
proportional to $\delta Z_\pi$ in the last line of eq.~(\ref{hampin}). For the last two diagrams
in fig.~\ref{ope2} we thus find:
\beq
h_{\rm 2N}^{(2)} = \left( \delta Z_\pi + \frac{8 M^2 d_{16}}{\krig{g}_A}-
\frac{4 M^2 d_{18}}{\krig{g}_A} \right)
\tilde \mathcal{A}_{\rm OPE}^{\rm NC}\,.
\eeq
Summing up all corrections to the OPE in the method of unitary transformation we end up 
with the same expression (\ref{OPEren}) as found from covariant perturbation theory using the 
S--matrix methods.

In summary, we have shown how to perform renormalization in the method of unitary transformation.
We recover the same expressions for pion and nucleon mass shifts as well as for corresponding
$Z$--factors. We have further demonstrated
that the NLO OPE potential derived using the method of unitary transformation, 
where the unitary operator is chosen according to eq.~(\ref{5.9}), coincides with 
the off--the--energy shell extension of the S--matrix. The TPE contribution has 
already been considered in both approaches, see refs.\cite{KBW,EGM1,EGM2} and 
identical results have been reported as well. We do not need to reconsider the TPE contributions and 
the pion loop diagrams which renormalize the short--range terms since no pion tadpoles appear in those 
cases and the results of refs.\cite{EGM1,EGM2} can be adopted without any changes.

\subsection{Explicit expressions for the potential at NLO}
\def\theequation{\arabic{section}.\arabic{equation}}
\label{sec:NLO}

We now give the explicit expressions for the chiral
effective NN potential $V_{\rm NLO}$,
which we use for extrapolation in the pion mass:
\beq
V_{\rm NLO} = V^{\rm OPE} + V^{\rm TPE} + V^{\rm cont}\,,
\eeq
where
\beqa
\label{potfin}
V^{\rm OPE} &=& - \frac{1}{4} \frac{g_A^2}{F_\pi^2} \left( 1 + 2 \Delta
- \frac{4 \tilde M_\pi^2}{g_A} \bar d_{18} \right) 
\, \fet \tau_1 \cdot \fet \tau_2 \, \frac{(\vec \sigma_1 \cdot \vec q \,) 
( \vec \sigma_2 \cdot \vec q\,)}
{\vec q\, ^2 + \tilde M_\pi^2} ~, \\
\label{potfinTPE}
V^{\rm TPE} &=& - \frac{ \fet{\tau}_1 \cdot \fet{\tau}_2 }{384 \pi^2 
f_\pi^4}\,\biggl\{
L(q) \, \biggl[4\tilde M_\pi^2 (5g_A^4 - 4g_A^2 -1) + \vec q\, ^2 (23g_A^4 - 
10g_A^2 -1) 
+ \frac{48 g_A^4 \tilde M_\pi^4}{4 \tilde M_\pi^2 + \vec q\, ^2} \biggr] \nn
&& {} \mbox{\hskip 2 true cm} + \vec q\, ^2 \, \ln \frac{\tilde M_\pi}{M_\pi} 
\, (23g_A^4 - 10g_A^2 -1) \biggr\} \\
&&{} - \frac{3 g_A^4}{64 \pi^2 f_\pi^4} \,\left( L(q) + 
\ln \frac{\tilde M_\pi}{M_\pi} \right)\, \biggl\{
\vec{\sigma}_1 \cdot\vec{q}\,\vec{\sigma}_2\cdot\vec{q} - \vec q\,^2 \, 
\vec{\sigma}_1 \cdot\vec{\sigma}_2 \biggr\}~, \nn
\label{potfincont}
 V^{\rm cont} &=& \bar C_S + \bar C_T (\vec \sigma_1 \cdot \vec \sigma_2 ) 
+ \tilde M_\pi^2 \, \left( \bar D_S - \frac{3 g_A^2}{32 \pi^2 F_\pi^4} ( 8 F_\pi^2  C_T- 5 g_A^2 + 2) 
\ln \frac{\tilde M_\pi}{M_\pi} \right) \\
&& {}+ \tilde M_\pi^2 \left( \bar D_T - \frac{3 g_A^2}{64 \pi^2 F_\pi^4} ( 16 F_\pi^2  C_T- 5 g_A^2 + 2) 
\ln \frac{\tilde M_\pi}{M_\pi} \right)\,
(\vec \sigma_1 \cdot \vec \sigma_2 ) \nn
&& {} + C_1 {\vec q \,}^2 
+ C_2 {\vec k \,}^2 + ( C_3 {\vec q \, }^2 + C_4 {\vec k \,}^2 ) 
( \vec \sigma_1 \cdot \vec \sigma_2 ) \nn
&& {} + i C_5 \frac{ \vec \sigma_1 + \vec \sigma_2}{2} 
\cdot ( \vec k \times \vec q \, ) 
+ C_6 ( \vec q \cdot \vec \sigma_1 ) 
( \vec q \cdot \vec \sigma_2 ) + C_7 ( \vec k \cdot \vec \sigma_1 )
( \vec k \cdot \vec \sigma_2 ) \,,
\nonumber
\eeqa
with $g_A$ and $F_\pi$ the physical values of the nucleon axial coupling and 
pion decay constant, respectively.
Here and in what follows  we denote the value of the pion mass by $\tilde M_\pi$ 
in order to distinguish it from the physical one denoted by $M_\pi$.
Further, 
\beq
L(q) \equiv L(| \vec q \,|) 
= \frac{\sqrt{4 \tilde M_\pi^2 + \vec q\, ^2}}{|\vec q \,|} \, 
\ln \frac{ \sqrt{4 \tilde M_\pi^2 + \vec q\, ^2}
+ | \vec q \, |}{2 \tilde M_\pi}\, ,
\eeq
and $\Delta$ represents the relative shift in the ratio $g_A/F_\pi$ compared to its 
physical value:
\beqa
\label{deltaCL}
\Delta &\equiv& \frac{\left(g_A/F_\pi\right)_{\tilde M_\pi}-\left(g_A/F_\pi\right)_{M_\pi}}{
\left(g_A/F_\pi\right)_{M_\pi}} \\
&=& \left( \frac{g_A^2 }{16 \pi^2 F_\pi^2} - \frac{4 }{g_A}
\bar{d}_{16} + \frac{1}{16 \pi^2 F_\pi^2} \bar{l}_4 \right) (M_\pi^2 - \tilde M_\pi^2) - 
\frac{g_A^2 \tilde M_\pi^2}{4 \pi^2 F_\pi^2} \ln \frac{\tilde M_\pi}{M_\pi} \, . \\
\nonumber
\eeqa
Note that in the TPEP we only take into account 
the explicit $M_\pi$--dependence and use the 
physical values for $g_A$ and $F_\pi$. This is perfectly sufficient at NLO since any 
shift in $g_A$ and $F_\pi$ for a different value of $M_\pi$ in the TPE 
is a N$^4$LO effect. 
We have expressed the potential in eqs.~(\ref{potfin})--(\ref{potfincont}) in such  a way that it coincides 
for  $\tilde M_\pi = M_\pi$ with the one given in \cite{EGM2}.\footnote{The term in eq.~(\ref{potfin}) 
which breaks the Goldberger--Treiman relation and is proportional to $\bar d_{18}$ 
is not shown explicitly in ref.~\cite{EGM2}.}
The constants 
$\bar C_{S,T}$ and $\bar D_{S,T}$ are related to the $C_{S,T}$
from \cite{EGM2} via
\beq
C_{S,T} = \bar C_{S,T} + M_\pi^2 \bar D_{S,T}\,.
\eeq
Note further that the short--range terms of the type $\tilde M_\pi^2 \ln \tilde M_\pi$
in eq.~(\ref{potfincont}) result from the two--pion exchange as well as from the renormalization 
of the leading--order contact forces by pion loops. 
It is important to stress that renormalization of the LECs  
$C_S$, $C_T$, $C_{1, \ldots 7}$ due to pion loops does not depend on the pion mass and 
thus is of no relevance for this work. 

We will now briefly discuss an important issue related to the renormalization of the 
NN potential. We have performed the calculation of the renormalized OPE in the previous section
without specifying a regularization scheme and gave only general (unregularized) expressions.
One might worry about the dependence of the results on the regularization scheme. 
Let us consider as a simple example  the nucleon mass shift in eq.~(\ref{nms}).
Applying dimensional regularization to the divergent integral $J_{12}$ one ends up with the 
finite shift given by eq.~(\ref{nmsDR}). The bare nucleon mass $\krig{m}$ in that case is finite and
the constant $c_1$ is not renormalized. Similarly, some other bare quantities like
$M$, $F$, $\krig{g}_A$ are finite. The situation might change if a different
regularization scheme is applied.
For example, performing a momentum cut--off regularization of $J_{12}$ in   eq.~(\ref{nms}) one finds:
\beq
\label{nmscut}
\delta m_N= - 4 c_1 M_\pi^2 - \frac{3 \krig{g}_A^2}{8 F^2} J_{12} = 
- 4 c_1 M_\pi^2 - \frac{3 \krig{g}_A^2}{8 F^2} \left( \alpha \Lambda^3 + \beta M_\pi^2 \Lambda + 
\frac{M_\pi^3}{4 \pi} + \mathcal{O} (\Lambda^{-1}) \right)\,,
\eeq
where $\Lambda$ is the momentum cut--off and the coefficients $\alpha$ and 
$\beta$ depend on the precise choice of the regulator. Taking the limit $\Lambda \rightarrow 
\infty$ and performing a redefinition of the constants $\krig{m}$, $c_1$ in order to absorb
the terms proportional to $\Lambda^3$ and $\Lambda$,
we obtain the same finite shift $- 3 \krig{g}_A^2 M_\pi^3/(32 \pi F^2)$ as in the case 
of dimensional regularization. The difference to the previously discussed case is 
that the bare nucleon mass and the bare 
constant $c_1$, which do not correspond to observable quantities, are now infinite. Also other bare 
parameters are infinite if the cut--off regularization is used, see e.g. refs.~\cite{HD98,Burga98}.   
It is, however, crucial to understand that, say, the 
shift in the nucleon mass from its observed value to the one in the chiral limit 
is a well defined and finite 
quantity and is not affected by a specific choice of the regularization scheme. The same  
holds true for $F_\pi$ and $g_A$. Therefore, we are free to adopt the results for the corresponding 
shifts in these constants 
found in CHPT analyses of various processes and based upon the dimensional regularization
(i.e.~we can use the values for the LECs in eq.~(\ref{deltaCL})). We will
specify the numerical values of the appropriate LECs in the next section.

To close this section let us make a comment on calculating various observables based upon the 
effective potential introduced above. 
The NN phase shifts are obtained using the T--matrix method.
The corresponding partial--wave projected 
Lippmann--Schwinger equation  reads:
\beq
\label{LSeq}
T_{l,\, l'}^{s  j} ( p \, ', \,  p\,) =
V_{l,\, l'}^{s  j} ( p \, ', \,  p\,) + \sum_{l''} \int \,
\frac{d^3 q}{(2 \pi)^3} V_{l,\, l''}^{s  j} ( p \, ', \,  q\,)
\frac{m_N}{ p\, ^2 -  q \, ^2 + i \epsilon} 
T_{l'',\, l'}^{s  j} ( q , \,  p\,)\,.
\eeq
Notice that the nucleon mass enters the expression for the two--nucleon propagator in the 
above equation. Thus apart from the explicit and implicit $\tilde M_\pi$--dependence 
of the NN force, which has been discussed above, one should, in principle, take into account 
the $\tilde M_\pi$--dependence of $m_N$ in the Lippmann--Schwinger equation. 
Since the nucleon mass $m_N$ is treated in the Weinberg power counting formally 
as a much larger scale compared to $\lambda_\chi$ (i.e. $m_N \sim \Lambda_\chi^2 /Q$), 
the shift in the nucleon mass gives the correction to the T--matrix which is suppressed by 3 powers 
of the small momentum scale and thus contribute at NNLO:
\beq
m_N = m_N^{\rm CL}  + 4 c_1^{r} M_\pi^2 + \ldots = m_N^{\rm CL} \left( 1 + \frac{4 c_1 M_\pi^2}{m_N^{\rm CL}} + 
\ldots \right) = m_N^{\rm CL} + \mathcal{O}
\left( \frac{Q^3}{\Lambda_\chi^3} \right)\,,
\eeq
where $m^{\rm CL}$ refers to the nucleon mass in the chiral limit ($=\krig{m}$ if dimensional regularization 
is applied). 
We have checked the above estimation numerically and found indeed a significantly smaller effect 
in the deuteron binding energy (in the chiral limit)
from the nucleon mass shift compared to various contributions at NLO. 
It goes without saying that the  potential $V (\vec p \, ', \, \vec p \, )$ is multiplied by the regulating 
functions $f_{\rm R} (| \vec p \, |)$, $f_{\rm R} (| \vec p \, ' |)$ 
in order to cut off the large momentum components in the 
Lippmann--Schwinger equation (\ref{LSeq}), which cannot be treated properly in effective field theory. 
We used here the same exponential function $f_R (| \vec p \, |) = \exp [ -\vec p \, ^4/\Lambda^4]$
as in ref.~\cite{3Nno3NF} and vary the cut--off $\Lambda$ in the range from 500 to 600 MeV.

\section{Results}
\def\theequation{\arabic{section}.\arabic{equation}}
\setcounter{equation}{0}
\label{sec:res}

\subsection{Chiral input and contact terms}
\def\theequation{\arabic{section}.\arabic{equation}}
\setcounter{equation}{0}
\label{sec:para}

For the OPE contribution in eq.~(\ref{potfin}) both explicit and implicit 
dependences on the pion mass are known, so that a parameter--free extrapolation 
to values $\tilde M_\pi$ away from the observed one is possible.

In what follows, we use $g_A=1.26$, $F_\pi=92.4$~MeV.
The constant $\bar l_4$ is fixed from the scalar radius 
of the pion and the corresponding 
value is $\bar l_4 = 4.3$  \cite{GL84}. The LEC $\bar d_{16}$ has recently been 
determined from the process $\pi N \rightarrow \pi \pi N$ \cite{Fet00_2}.  
In the following we will use the updated value from ref.~\cite{FetTH}. 
To be more specific, we average the values obtained in the 3 different fits in 
the last reference. This leads to $\bar d_{16} = -1.23$ GeV$^{-2}$.
Note that although all values of $\bar d_{16}$ from  \cite{FetTH} have the 
same sign (negative) and are of a similar magnitude,
the uncertainty in the determination of this constant remains quite large.
We come back to this at the end of this section.
The constant $\bar d_{18}$ is fixed from the observed value of the Goldberger--Treiman 
discrepancy. 
Using the values of $g_{\pi N}$ extracted from 
$\pi N$ phase shift analysis \cite{Em98}, we get the following 
value for the LEC $\bar d_{18}$: $\bar d_{18} = -0.97$ GeV$^{-2}$.
Thus we obtain for the ratio $g_{\pi N}/m_N$, which is nothing but the 
strength of the OPE and is related to  $g_A$ and $F_\pi$ via 
(at the given order in  the chiral expansion)  
\beq
\label{gpin}
\frac{g_{\pi N}}{m_N} = \frac{g_A}{F_\pi} 
\left( 1 - \frac{ 2 M_\pi^2}{g_A} \bar d_{18} \right)\,,
\eeq
the following value in the chiral limit 
\beq
\left( \frac{g_{\pi N}}{m_N} \right)_{\rm CL}  \sim 15.0 \, \mbox{GeV}^{-1}\,.
\eeq
Comparing this value with the physical one of 13.2 GeV$^{-1}$, 
we conclude that the OPE  becomes {\em stronger} in the chiral limit.

We now turn our attention to the remaining contribution in the 
NN potential (\ref{potfin}).
The TPE part is parameter--free and the values of the LECs $C_{1,\ldots,7}$ 
as well as combinations $\bar C_{S,T} + M_\pi^2 \bar D_{S,T}$ have
already been fixed from the fit to low--energy data in the NN S-- and P--waves. 
Unfortunately, the LECs $\bar D_{S,T}$, which contribute to S--wave projected 
potential, are not known at the moment. 
Ideally, they should be fixed from the $NN\pi$ system in processes 
like e.g. pion--deuteron
scattering. This has not yet been done. In order to proceed 
further we assume natural values for  these constants, i.e.:
\beq
\bar D_{S,T} = \frac{\alpha_{S,T}}{F_\pi^2 \Lambda_\chi^2}\, ,  
\quad \mbox{where} \quad \alpha_{S,T} \sim 1\,,
\eeq
and $\Lambda_\chi \simeq 1$ GeV.
In  our analysis in \cite{EGMres} we have shown that {\bf all values} 
of the dimensionless coefficients $\alpha$
related to the contact terms lie at NLO in the range $-2.1 \ldots 3.2$ 
for all  cut--offs employed. Thus making the conservative estimate 
\beq
\label{estimD}
-3.0 < \alpha_{S,T} < 3.0\,,
\eeq
we are rather confident to be on the safe side. 
Note that this corresponds to the following range 
for the partial--wave projected values:
\beq
\label{Destim}
 -\frac{48  \pi}{F_\pi^2 \Lambda_\chi^2} < \bar D_{^1S_0} < \frac{48  \pi}{F_\pi^2 \Lambda_\chi^2}\,,
\quad \quad \quad
 -\frac{24  \pi}{F_\pi^2 \Lambda_\chi^2} < \bar D_{^3S_1} < \frac{24  \pi}{F_\pi^2 \Lambda_\chi^2}\,.
\eeq
It is worth mentioning that the term $\tilde M_\pi^2  
\bar D_{T} (\vec \sigma_1 \cdot \vec \sigma_2)$
breaks Wigner symmetry. The value of $\bar D_T$ 
expected from eq.~(\ref{Destim}) leads
to a Wigner symmetry breaking effect that is comparable in size with 
the one from the observed nonvanishing (leading order) 
value of $C_T$ \cite{EGMres} (i.e. $M_\pi^2 \bar D_T \sim C_T$), 
so that the possibility of exact Wigner symmetry in the chiral limit
exists. In that case, Wigner symmetry breaking would be related to
the explicit chiral symmetry breaking of QCD.
Last but not least, 
we use here the values 
of the LEC $C_{S,T}$, $C_{1,\ldots,7}$ 
from \cite{EGM2}.\footnote{In our earliar work \cite{EGM2} we used 
a value for the pion--nucleon coupling $g_{\pi N}$, 
which is slightly smaller than the commonly accepted 
one $g_{\pi N} = 13.1 \ldots 13.4$. We have refitted the contact 
terms $C_{S,T}$, $C_{1,\ldots,7}$ using 
the value for $g_{\pi N}$ according to eq.~(\ref{gpin}). 
This leads to minor changes in the values 
of the LECs related to contact terms and does not visibly change observables.}

Before showing our results for the phase shifts and other quantities,
we would like to make a comment on
the uncertainty of our extrapolation in $\tilde{M_\pi}$. As already pointed out, 
the main source of uncertainty is related to the unknown values of the LECs 
$D_{S,T}$, which enter the expression for contact interactions at NLO, see 
eq.~(\ref{potfincont}). Although the LECs $\bar l_4$, $\bar d_{16, 18}$ are 
known from CHPT analyses of various processes and thus the complete 
$\tilde M_\pi$--dependence of the ratio $g_{\pi N}/m_N$ is fixed at NLO, some 
uncertainty still remains due to the uncertainty in the determination of 
the abovementioned LECs. While the constant $l_4$ is known with relative
small error bars, so that the corresponding uncertainty needs not to be
discussed here, the LECs  $\bar d_{16, 18}$ are not known very precisely. 
Using the values of $g_{\pi N}$ extracted from 
three different $\pi N$ phase shift analyses \cite{Ka85,Em98,Sp98} 
one gets the values $\bar d_{18} = -1.54$ GeV$^{-2}$, $\bar d_{18} = -0.97$ GeV$^{-2}$
and $\bar d_{18} = -0.84$ GeV$^{-2}$, respectively. From eq.~(\ref{potfin}) one
sees that $\bar d_{18}$ does not contribute to the OPE in the chiral limit. 
Thus at first sight our predictions in the chiral limit seem not to be  
affected by the uncertainty in this LEC. In fact, this is not quite true, since 
changing the value of $\bar d_{18}$ leads to a modification of the strength of the OPE at the 
physical point ($\tilde M_\pi = M_\pi$). Therefore, one should, in principle,
refit at the same time the LECs $C_{S,T}$, $C_{1, \ldots , 7}$ in order to 
describe the corresponding phase shifts. Modifying  the LECs related to 
the contact interactions would change observables even in the chiral limit.    
However, we have checked that the corresponding effects in observables are 
small and will not discuss this issue in what follows. The uncertainty in the 
LEC $\bar d_{16}$ turns out to be much more important for our analysis. 
Contrary to the previously discussed case with the LEC $\bar d_{18}$, changing 
$\bar d_{16}$ result in changes of the strength of the 
OPE in the chiral limit (and, of course, 
for other values of $\tilde M_\pi \neq M_\pi$) but does not affect it at the physical
point where $\tilde M_\pi = M_\pi$. In what follows we will demonstrate 
how the uncertainty in   $\bar d_{16}$ shows up in various observables.

\subsection{Phase shifts}
\def\theequation{\arabic{section}.\arabic{equation}}
\setcounter{equation}{0}

\begin{figure}[htb]
%\vspace{1.2cm}
\centerline{
\psfig{file=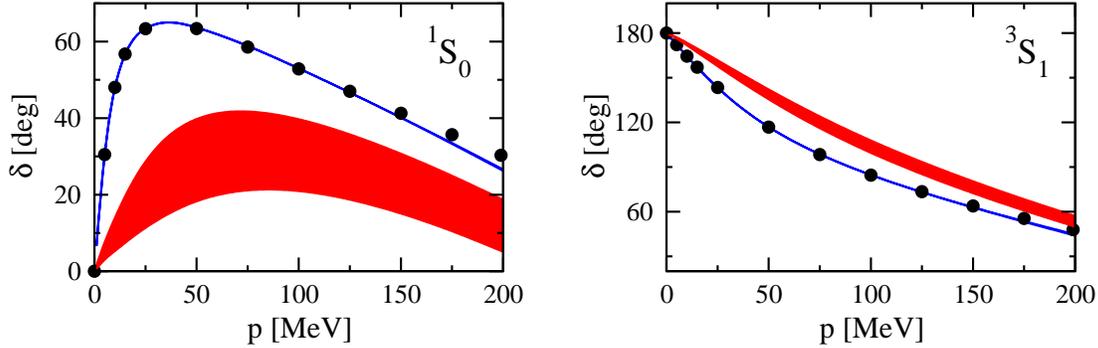,width=16cm}}
\vspace{0.3cm}
\centerline{\parbox{14cm}{
\caption[fig4]{\label{sw} 
S--wave phase shifts as functions of the momentum in the c.m. system. 
Dark shaded bands correspond to our NLO predictions in the chiral limit, light shaded bands are NLO results 
for physical value of $M_\pi$, filled circles are Nijmegen phase shifts \cite{NPSA}.
}}}
\vspace{0.7cm}
\end{figure}

\begin{figure}[htb]
%\vspace{1.2cm}
\centerline{
\psfig{file=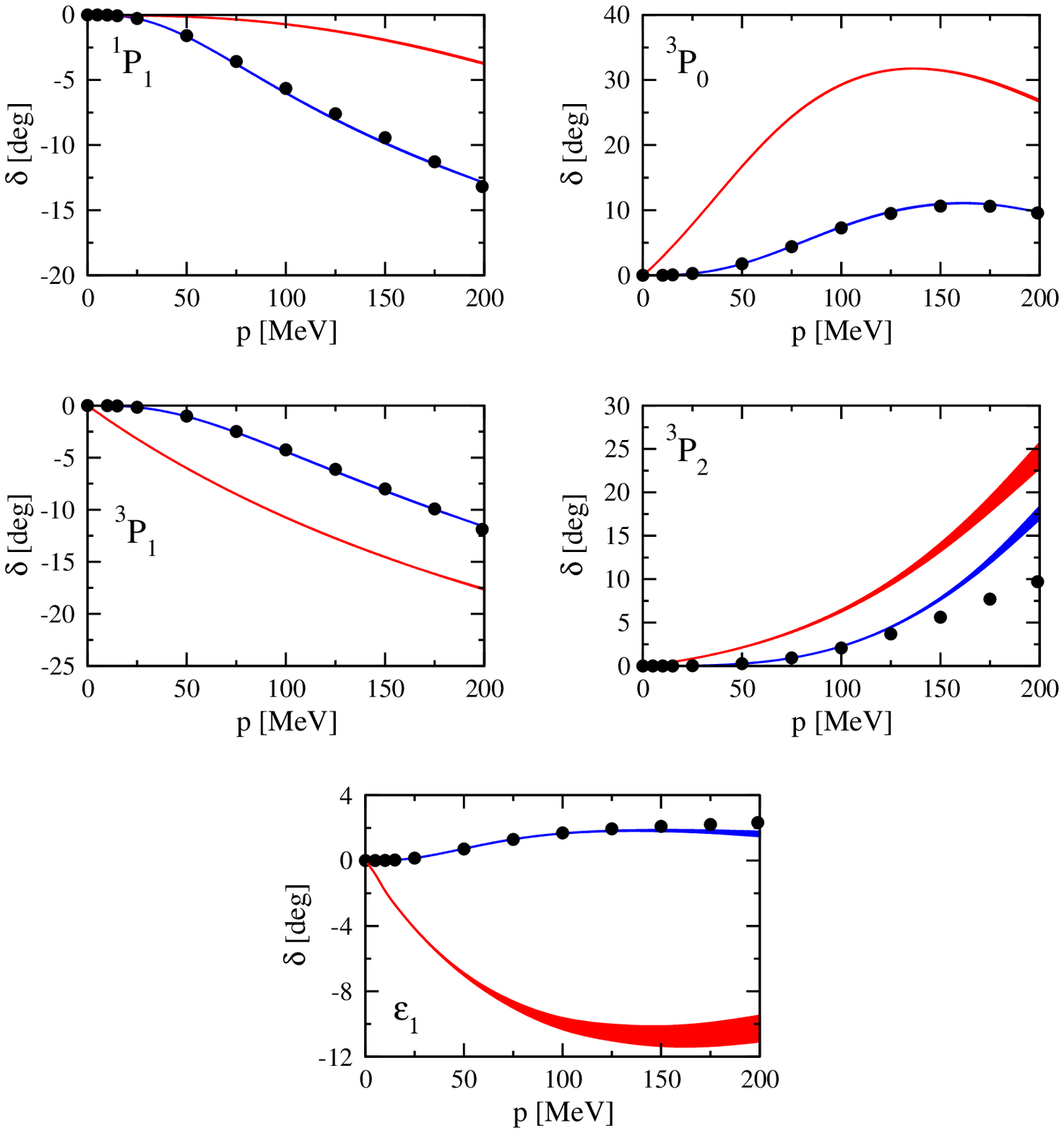,width=16cm}}
\vspace{0.3cm}
\centerline{\parbox{14cm}{
\caption[fig4]{\label{pw} 
P--wave phase shifts. For notations see fig.~\ref{sw}.
}}}
\vspace{0.7cm}
\end{figure}

\begin{figure}[htb]
%\vspace{1.2cm}
\centerline{
\psfig{file=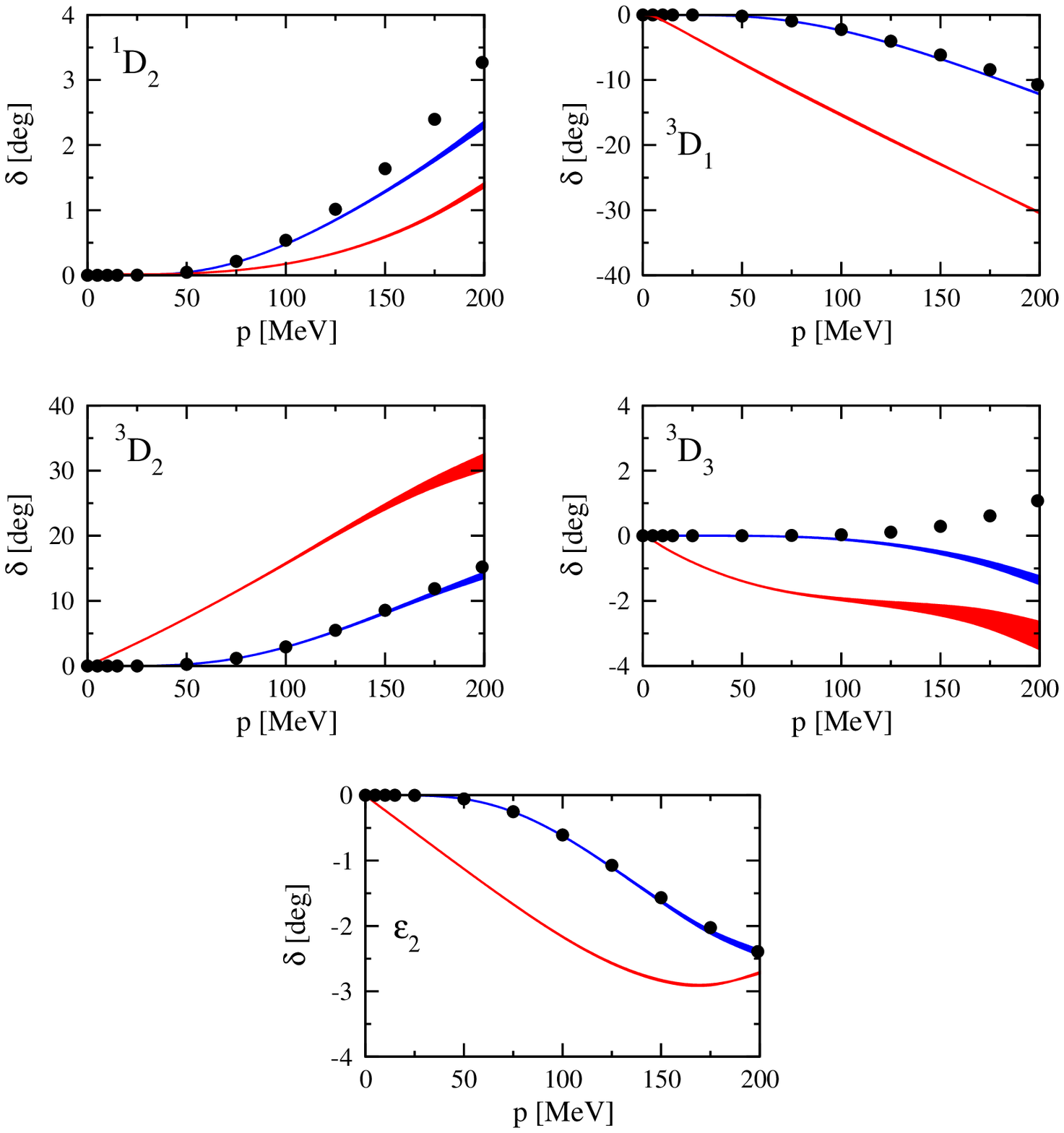,width=16cm}}
\vspace{0.3cm}
\centerline{\parbox{14cm}{
\caption[fig4]{\label{dw} 
D--wave phase shifts.  For notations see fig.~\ref{sw}.
}}}
\vspace{0.7cm}
\end{figure}

\begin{figure}[htb]
\centerline{
\psfig{file=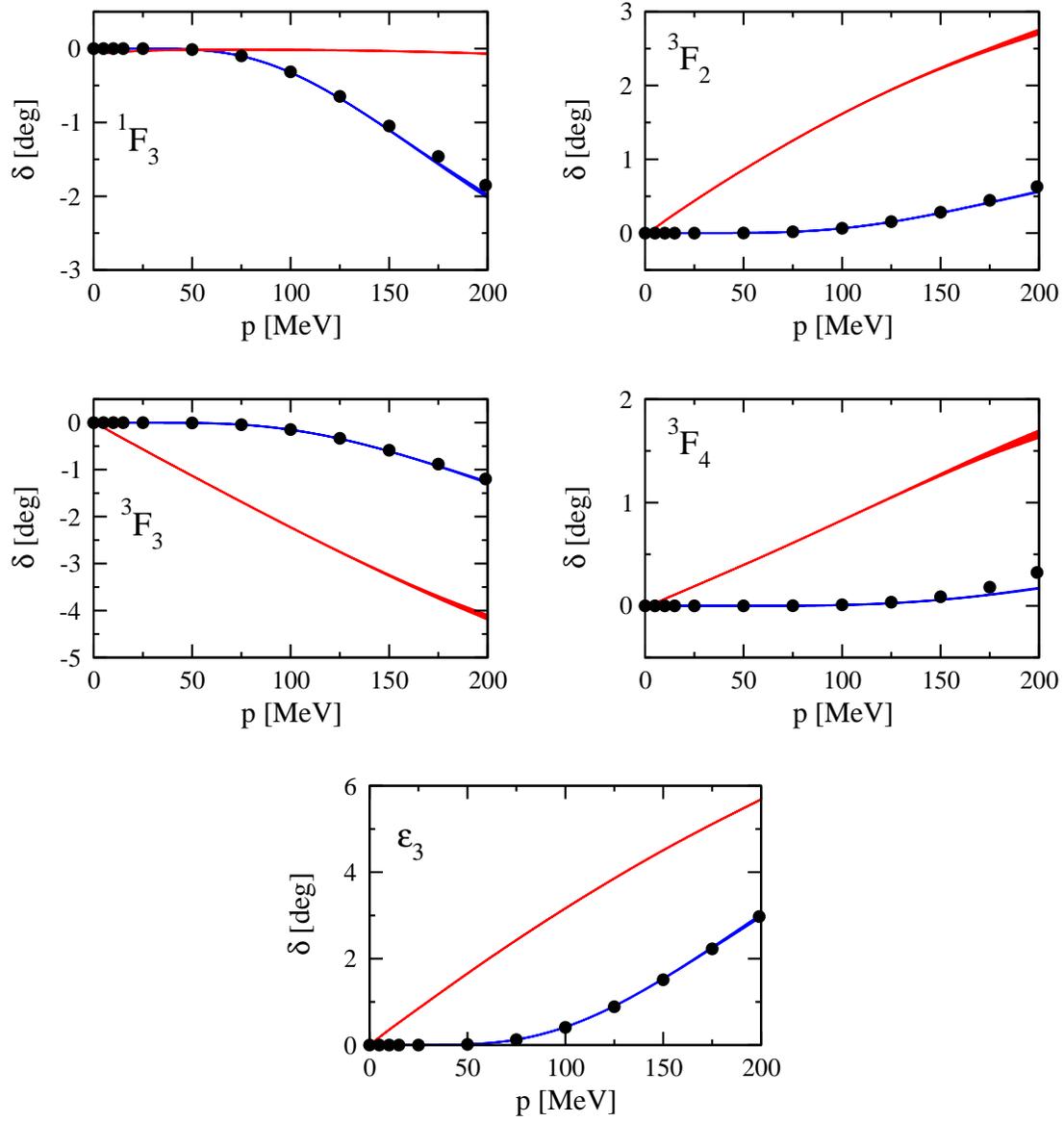,width=16cm}}
\vspace{0.3cm}
\centerline{\parbox{14cm}{
\caption[fig4]{\label{fw} 
F--wave phase shifts.  For notations see fig.~\ref{sw}.
}}}
\vspace{0.7cm}
\end{figure}

After the remarks of the preceding section we are now in the position to present results.
In figs.~\ref{sw} to \ref{fw}  we show our predictions for various 
phase shifts in the chiral limit compared to their values in the real world. 
In order to estimate the uncertainty due to missing higher order terms we proceed in a usual way
and vary the cut--off between $500$ and $600$ MeV, as it has been done in \cite{3Nno3NF}. 
The only exception to this are the $^1S_0$ and $^3S_1 {}- {}^3D_1$ channels in the chiral limit,
where the dark shaded bands in figs.\ref{sw},\ref{pw},\ref{dw} are due to the uncertainty in the 
values of $\bar D_{S,T}$. Varying the cut--off in that case is not necessary, since the 
corresponding uncertainty is much smaller than the one due to  unknown $\bar D_{S,T}$. 
For these channels we will take the cut--off $\Lambda= 560$ MeV.

Both S--wave phase shifts in fig.~\ref{sw} 
show a qualitatively similar behaviour to what is observed in the real world.
In the $^1S_0$ channel the interaction in the chiral limit gets weaker and the phase
shift reaches in the maximum about $30 - 50 \%$  of the observed value. The situation in
the $^3S_1$ channel is just opposite but the effect is smaller in magnitude.
Note that the band for the  $^1S_0$ phase shift 
is much wider, partially due to the larger uncertainty in $\bar D_{^1S_0}$ compared to the one 
in $\bar D_{^3S_1}$.

Before considering higher partial waves some comments are in order. 
First of all we note that the form of the OPE in the chiral limit
\beq
\label{opecl}
(V^{\rm OPE})_{\rm CL} \propto \frac{(\vec \sigma_1 \cdot \vec q \, ) 
(\vec \sigma_2 \cdot \vec q \, )}{\vec q \, ^2}
\eeq
leads to significant scattering at vanishing momenta for all partial waves, as pointed out in 
by Bulgac et al. \cite{Bul97}. 
It is amusing to note that although the interaction between pions and nucleons vanishes in the 
chiral limit at vanishing momenta, the interaction between nucleons via the exchange of pions, 
becomes,
on the contrary, stronger as a consequence of 
the increased range of the interaction for $\tilde M_\pi 
\rightarrow 0$.\footnote{The OPEP behaves at large distances  in the chiral limit as $1/r^3$.}
As a consequence of the vanishing pion mass  the effective range expansion 
\beq
k^{2 l +1} \cot \delta_l (k) = - \frac{1}{a_l} + r_l \frac{k^2}{2} + v_l^{2} k^4 + \ldots\,,
\eeq
where $k$ is the c.m. momentum and $l$ the angular momentum, 
does not exist any more. The partial wave amplitude $A_l (E)$, $E=k^2/m_N$, 
describing elastic NN scattering has a left--hand cut starting at the 
branch point $E=-\tilde M_\pi^2/(4 m_N)$, 
which gives a maximal radius of convergence of the effective range expansion. Thus the 
domain of validity of the effective range expansion ceases to exist 
for $\tilde M_\pi \rightarrow 0$. 

Projecting the OPE in the chiral limit, eq.~(\ref{opecl}), onto the states with 
definite $l$, $s$ and $j$ 
we note that all spin--singlet matrix elements vanish. 
Further, on--the--energy shell matrix elements in 
all other channels do not depend on the momenta. As a consequence, 
one expects that the phase shifts 
at low energy in the high partial waves, in which the amplitude is essentially given 
by the Born term and 
strongly dominated by the OPEP, behave like $\delta (k) \propto k$.

Let us now comment on the P--waves which are, 
as already pointed out in the introduction, of a particular interest,
because of the possible existence of  bound states  due to the 
strong OPE  in the chiral limit. First,
we stress that we are able to perform an accurate extrapolation 
in the pion mass for P-- and higher partial waves since the complete 
$\tilde M_\pi$--dependence is known.
Our findings confirm the conclusions of \cite{Bul97}
that no bound states exist in the P--waves. The phase shift in the $^3P_0$ channel, which is 
most sensitive to the OPE and thus is an ideal candidate for the appearance of a bound state, 
is strongly enhanced compared to the physically relevant case and reaches a maximum 
of about $32^\circ$.
The $^1P_1$ phase shift becomes small and is completely given  in the chiral limit 
by the TPEP and the NLO contact term. 
Remarkably, one observes strong changes in the mixing angle $\epsilon_1$, which
is known to be an observable that is very sensitive to small changes in certain
parameters. It is comforting to note that 
the bands do not get significantly wider compared to the case with the physical value of the
pion mass,  $\tilde M_\pi = M_\pi$. This is a clear indication of the  consistency and 
good quality of the extrapolation to $\tilde M_\pi =0$, since the 
cut--off independence of observables in the chiral limit is achieved by ``running''  
of the LECs $C_i (\Lambda)$ fixed at the physical value $\tilde M_\pi = M_\pi$.

The situation with the D-- and F-- waves turns out to be similar to the case of the P--waves.
The phases in most channels (apart from the spin--singlet ones) already follow their 
asymptotical behaviour discussed above, i.e. grow linearly with momentum $k$. 
Our last comment in this section is that the partial wave decomposition becomes 
not a reasonable 
(from the practical point of view) tool for calculating observables (full scattering amplitude) 
if $M_\pi \rightarrow 0$,
since the expansion in the partial waves converges slowly in the presence of the long--range 
interaction $\propto 1/r^3$.

\subsection{Deuteron binding energy}
\def\theequation{\arabic{section}.\arabic{equation}}
\setcounter{equation}{0}

\begin{figure}[htb]
\centerline{
\psfig{file=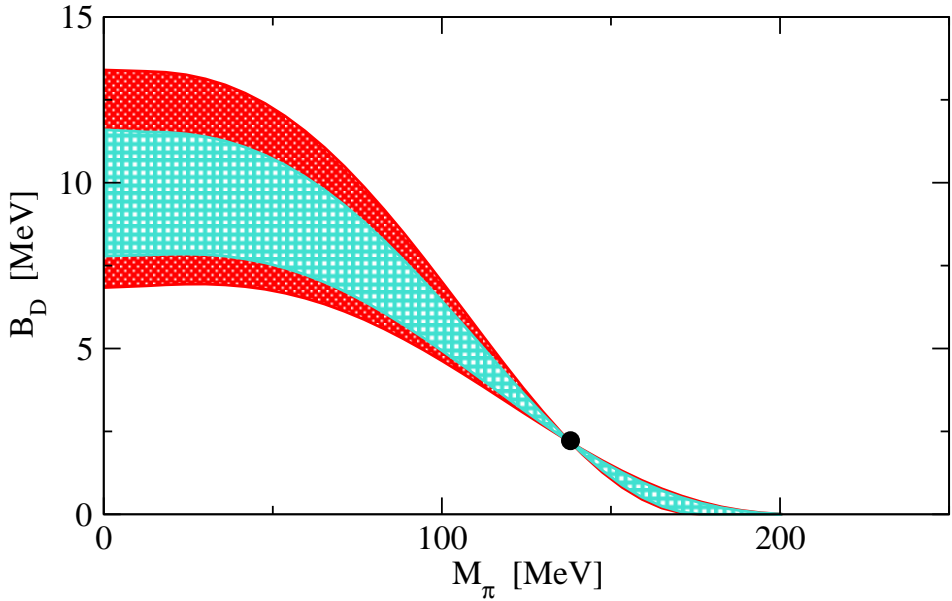,width=14cm}}
\vspace{0.3cm}
\centerline{\parbox{14cm}{
\caption[fig4]{\label{deut} 
Deuteron binding energy as a function of the pion mass. The shaded areas correspond 
to allowed values. 
The light shaded band corresponds to our main result with  
$\bar d_{16} = -1.23$ GeV$^{-2}$ and the uncertainty due to the unknown LECs $\bar D_{S,T}$.
The dark shaded band gives the uncertainty if, in addition to variation of $\bar D_{S,T}$, 
the LEC $\bar d_{16}$ is varied in the range from 
$\bar d_{16} = -0.91$ GeV$^{-2}$ to $\bar d_{16} = -1.76$ GeV$^{-2}$ given in \cite{FetTH}.
The heavy dot shows the binding energy for the physical value of the 
pion mass. 
}}}
\vspace{0.7cm}
\end{figure}

We now move on to  discuss the deuteron binding energy $B_{\rm D}$, which we consider
to be the most interesting observable with respect to its $\tilde M_\pi$--dependence,
see fig.~\ref{deut}.
As already pointed out in the introduction, the situation concerning the
chiral limit behaviour of $B_{\rm D}$ 
according to the previous analyses of Bulgac et al.~\cite{Bul97} and 
Beane et al.~\cite{Beane01,BeaneNEW} can not be considered as settled.
In fact, the value of $B_{\rm D}$ is rather sensitive to the assumptions
and approximations made as well as to certain parameters, like the
pion-nucleon coupling constant. 
According to our  complete NLO analysis, we arrive at a unique result: the 
deuteron  is {\em stronger} bound in the chiral limit with the binding energy
\beq
\label{deutbind}
B_{\rm D}^{\rm CL} =  9.6 \pm 1.9 {{+ 1.8} \atop  
{-1.0}}
\,\, \mbox{MeV}\,.
\eeq
Here, the first  error refers to the uncertainty in the value of $\bar D_{^3S_1}$, where
the LEC $\bar d_{16}$ is set to the average value  $\bar d_{16}=-1.23$ GeV$^{-2}$ from \cite{FetTH}.
The second  error shows the additional uncertainty when $\bar d_{16}$ is varied
in the range given in \cite{FetTH}.
We decided not to add these uncertainties but rather give them  
separately since they are not completely uncorrelated. 
The value of $B_{\rm D}^{\rm CL}$ is in remarkable agreement with the natural 
value one would expect to arise in QCD of about $F_\pi^2/m \sim
10$~MeV, if one (naively) assumes for the binding energy
to be of the order of the kinetic energy with the relevant momentum scale
being $p \sim F_\pi$.\footnote{More precisely, one should use in this
  formula $F$ and $\krig{m}$, but for an approximate estimate the differences
  to the physical values can be ignored.} 
\noindent

We also performed extrapolation for larger values of $\tilde M_\pi$, which might be
of interest for later calculations within lattice QCD. 
According to our analysis the deuteron does not exist anymore for 
$\tilde M_\pi \gtrsim 200$ MeV.
Note that the behaviour of $B_{\rm D}$ is qualitatively similar to what has been 
found in \cite{Beane01} by taking into account only the explicit $\tilde M_\pi$--dependence.

\subsection{S--wave scattering lengths and matching with lattice QCD}
\def\theequation{\arabic{section}.\arabic{equation}}
\setcounter{equation}{0}

\begin{figure}[htb]
%\vspace{1.2cm}
\centerline{
\psfig{file=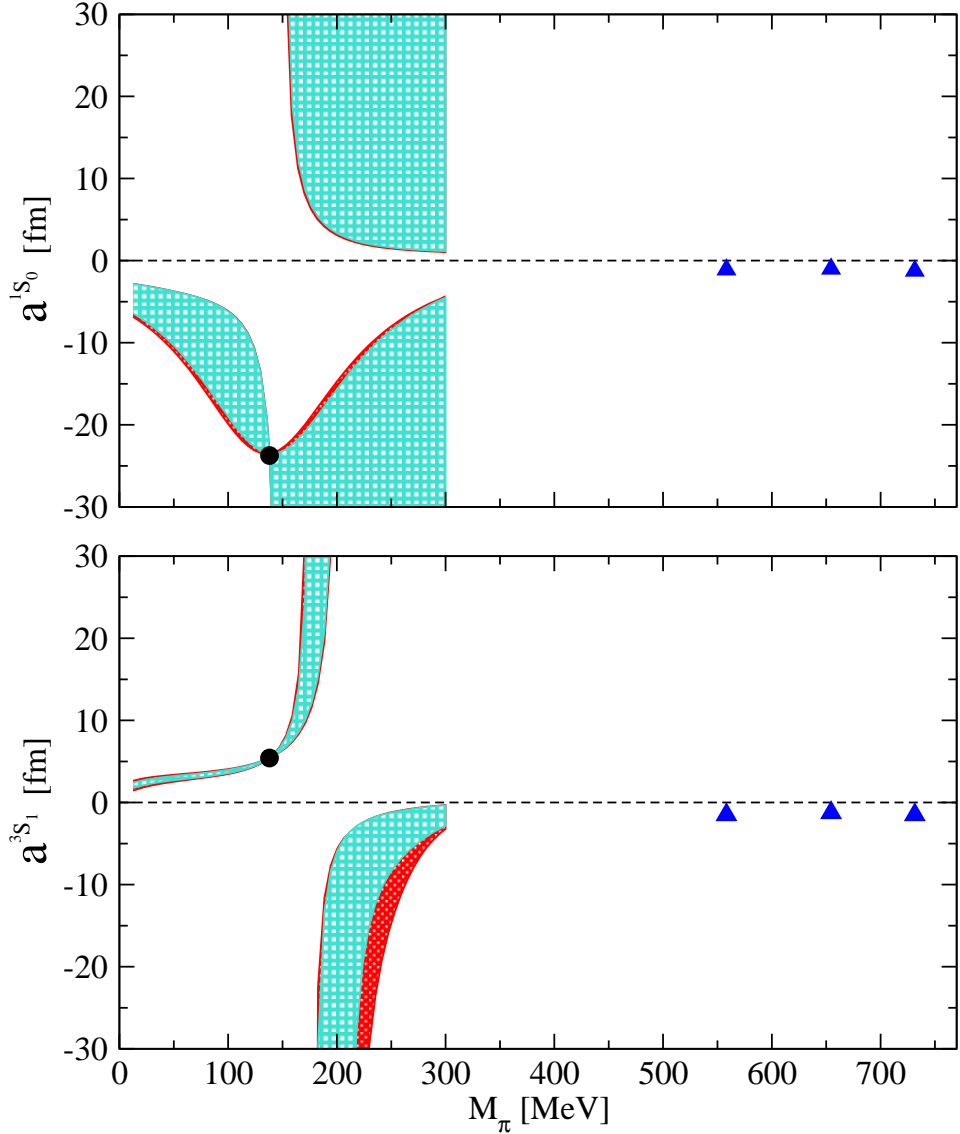,width=14cm}}
\vspace{0.3cm}
\centerline{\parbox{14cm}{
\caption[fig4]{\label{scattl} 
The S--wave scattering lengths as functions of $\tilde M_\pi$. 
The shaded areas represent the allowed values according to our analysis.
The light shaded band corresponds to our main result with  
$\bar d_{16} = -1.23$ GeV$^{-2}$ and the uncertainty due to the unknown LECs $\bar D_{S,T}$.
The dark shaded band gives the uncertainty if, in addition to variation of $\bar D_{S,T}$, 
the LEC $\bar d_{16}$ is varied in the range from 
$\bar d_{16} = -0.91$ GeV$^{-2}$ to $\bar d_{16} = -1.76$ GeV$^{-2}$ given in \cite{FetTH}.
The heavy dots corresponds to the values in the real world. The triangles 
refer to lattice QCD results from \cite{Fukug95}. 
}}}
\vspace{0.7cm}
\end{figure}

Our last topic is related to the  S--wave scattering
lengths $a^{^1S_0}$ and $a^{^3S_1}$  and the possibility of matching 
with the calculations using lattice QCD. 
In fig.~\ref{scattl} we show our predictions for the scattering lengths.
First of all, it is interesting to see that the uncertainty due to 
the unknown constant $\bar D_{^3S_1}$ does not show up strongly for $a^{^3S_1}$
for $\tilde M_\pi < M_\pi$, where we are able to obtain an accurate extrapolation.
In the chiral limit the scattering length $a^{^3S_1}_{\rm CL}$ is still 
positive and smaller in magnitude compared to 
the physically relevant case, which is in agreement with the larger value of the deuteron 
binding energy. Specifically, we get $a^{^3S_1}_{\rm CL} = 1.5 \pm 0.4
{{+ 0.2} \atop 
{-0.3}} \, $ fm, which is again
of natural size. Here the first uncertainty refers to variation in $\bar D_{S,T}$
while the second one to the additional variation in $\bar d_{16}$, as described above.
The scattering length changes its sign at $\tilde M_\pi \sim 200$ MeV and 
relaxes to natural values for larger $\tilde M_\pi$. Note also that the relative  
contribution of the OPE decreases with increasing pion mass and thus the uncertainty
due to the variation in $\bar{d}_{16}$ becomes very small. On the contrary, the
uncertainty in $\bar D_{^3S_1}$ is, of course, magnified as $\tilde{M}_\pi$ grows.

\begin{figure}[htb]
\centerline{
\psfig{file=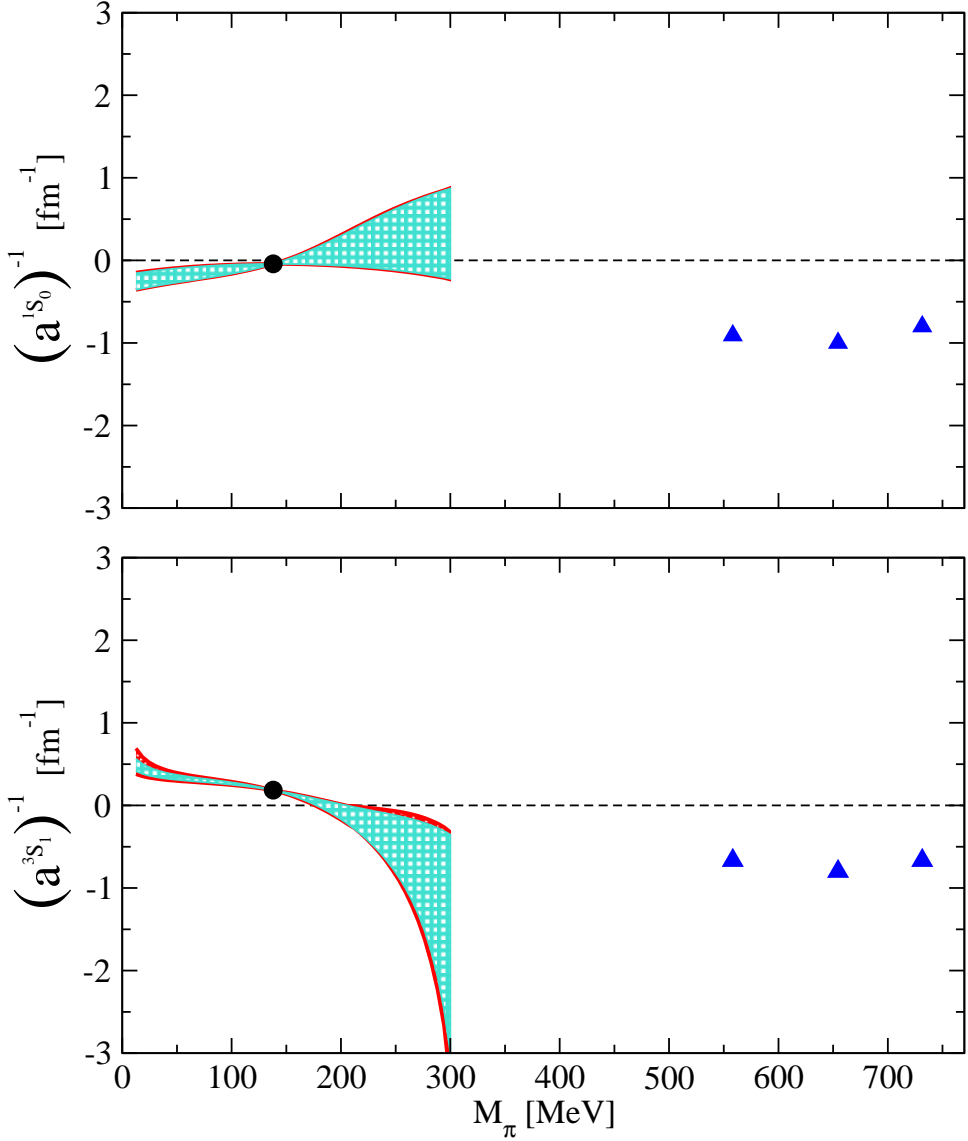,width=14cm}}
\vspace{0.3cm}
\centerline{\parbox{14cm}{
\caption[fig4]{\label{scattlinv} 
The S--wave inverse scattering lengths as functions of $\tilde M_\pi$. 
The shaded areas represent the allowed values according to our analysis.
The light shaded band corresponds to our main result with  
$\bar d_{16} = -1.23$ GeV$^{-2}$ and the uncertainty due to the unknown LECs $\bar D_{S,T}$.
The dark shaded band gives the uncertainty if, in addition to variation of $\bar D_{S,T}$, 
the LEC $\bar d_{16}$ is varied in the range from 
$\bar d_{16} = -0.91$ GeV$^{-2}$ to $\bar d_{16} = -1.76$ GeV$^{-2}$ given in \cite{FetTH}.
The heavy dots corresponds to the values in the real world. The triangles 
refer to lattice QCD results from \cite{Fukug95}. 
}}}
\vspace{0.7cm}
\end{figure}

In case of the $a^{^1S_0}$ the uncertainty in our extrapolation in the pion mass 
is visibly larger. This is consistent with the known fact that the OPE plays a less 
significant role in this channel, where the interaction is dominated by the shorter
range terms. Another reason is that an extreme fine tuning leading to the very large 
value for $a^{^1S_0}$ takes place in the 
physically relevant case with $\tilde M_\pi = M_\pi$, where all parameters are fixed.
The larger effect due to the uncertainty in the $\bar D_{^1S_0}$ thus has 
to be expected. Nevertheless, our results concerning this channel are also
quite interesting. In particular, 
it turns out that no bound state appears for  pion mass smaller than $M_\pi$, although 
in the real world for $\tilde M_\pi = M_\pi$ the virtual state in this channel is almost bound.
Around the chiral limit $a^{^1S_0}$ shows a qualitatively similar behaviour to 
$a^{^3S_1}$ and gets smaller in magnitude: $a^{^1S_0}_{\rm CL} = -4.1 \pm 1.6 
{{+ 0.0} \atop  
{-0.4}} \,$ fm,
which is still somewhat large compared to what would be expected from dimensional reasons
($1-2\,$fm). We are not able to obtain an accurate prediction for $a^{^1S_0}$ for  
$\tilde M_\pi > M_\pi$. In that case both scenarios with or without a bound state in this channel 
are possible. This uncertainty leads to a huge uncertainty with respect to the scattering
length, which becomes infinite in the presence of a zero--energy bound state.
This situation is depicted in the upper panel of fig.~\ref{scattl}. 
The different signs 
of the scattering length are due to the fact that the two--nucleon system might be 
bound or unbound in this channel.
It is more appropriate in such a case to look at the inverse scattering length,
which simply crosses zero when one encounters a zero energy bound state. 
We show the inverse scattering lengths
$1/a^{^1S_0}$ and $1/a^{^3S_1}$ in fig.~\ref{scattlinv}.
Unfortunately, it is not yet possible to compare our extrapolation of 
the scattering lengths with the results available from  the lattice
calculation. The latter ones are currently restricted to values of 
$\tilde M_\pi$ larger than 550 MeV, which is far beyond the domain of 
applicability of the chiral expansion. Actually, our extrapolation is 
not trustable anymore for $\tilde M_\pi \gtrsim 2 M_\pi$. For example, 
the relative shift $\Delta$ of the ratio $g_A/F_\pi$ defined in 
eq.~(\ref{deltaCL}) reaches $\sim 50 \%$
for $\tilde M_\pi \sim 250$ MeV.  Furthermore, the lattice calculation 
of ref.~\cite{Fukug95} is carried out in the quenched approximation, 
so that one, in principle, should perform extrapolation in the pion mass
in the quenched chiral perturbation theory, see \cite{BS1,BS2}
for more discussion.

\subsection{Comparison to earlier work}
\def\theequation{\arabic{section}.\arabic{equation}}
\setcounter{equation}{0}

It is interesting to compare our findings with the ones of
refs.~\cite{Beane01,BeaneNEW}, which are also based on effective field theory.
There it was pointed out that the (natural) values of the unknown LECs $\bar D_{S,T}$ 
exist, which make the deuteron both bound or unbound in the chiral limit and which strongly
change the S--wave scattering lengths. Thus no predictions for all these quantities 
could be made \cite{BeaneNEW}. The basic differences in our analyses can be summarized as follows:
\begin{enumerate}
\item
While we take the chiral limit value for the strength of the OPE, $g_A/F_\pi$, 
from the CHPT analysis of the $\pi \pi$,
$\pi N$ and $\pi \pi N$ systems, the authors of ref.~\cite{BeaneNEW} make use of the assumption
that terms of the type $M_\pi^2 \ln M_\pi/\Lambda$, where $M_\pi \ll \Lambda$ ($\Lambda \sim M_\rho$)
dominate the chiral expansion. This leads to a decrease of $g_A/F_\pi$ in the chiral limit,
while the CHPT based analysis gives an increased value. This difference seems to be the 
most important one.
\item
While we take into account the complete TPEP, only its chiral limit value is considered in 
\cite{BeaneNEW}. In addition, we also include renormalization of the short--range interactions,
which leads to the nontrivial logarithmic dependence on the pion mass.
\item
We regularize the Lippmann--Schwinger in the momentum space by a finite cut--off 
and do not require the low--energy phase shifts to be completely but only approximately cut--off 
independent. The remaining cut--off dependence can be eliminated by inclusion of higher order terms.
In \cite{BeaneNEW} the Schr\"odinger equation in the coordinate space (with the finite 
short--distance cut--off) is solved. The cut--off independence of the observables 
is achieved at the expense of introducing short--range interactions in the D--wave, which are
not present in the effective Lagrangian.
\end{enumerate}
Finally, we remark that as in ref.\cite{Bul97} we find no new bound states in the P-waves.

\section{Summary and conclusions}
\def\theequation{\arabic{section}.\arabic{equation}}
\setcounter{equation}{0}
\label{sec:summ}

In this paper we have calculated properties of the two--nucleon system in the chiral 
limit based on
a chiral effective field theory. The results of our investigation can be summarized as follows:
\begin{itemize}
\item
Based upon the modified Weinberg power counting (as explained in \cite{EGM1}) we have 
shown how to perform a complete renormalization within the method of unitary transformation,
including the mass and wave--function renormalization and how to deal with  tadpole graphs.
We found the renormalized expression for the OPEP within the projection formalism and 
demonstrated that
it agrees precisely with the off--the--energy shell extension of the OPE amplitude obtained
in the S--matrix method. 
\item
Based on the NLO potential, which includes the renormalized OPE, TPE contributions and 
contact terms, 
we have performed extrapolations in the pion mass 
(or, equivalently, quark mass) away from its 
physical value \footnote{It certainly would be interesting  to extend the
  calculation to NNLO. }. 
The corresponding LECs have been taken from an 
investigation of $\pi \pi$ \cite{GL84}, $\pi N$  \cite{Fet00,FetTH}
and $NN$ systems \cite{EGM2}.
The only uncertainties arise from the unknown couplings $\bar D_{\rm S,T}$ related to 
contact terms with one insertion  $\sim \tilde M_\pi^2$ and to the variation in the
pion--nucleon LEC $\bar{d}_{16}$ related to the Goldberger-Treiman discrepancy.
For the dimensionless coefficients that parameterize the LECs  $\bar D_{\rm S,T}$ and
which are expected to be of order one, we considered
values in the range from $-3$ to $3$. The variation in  $\bar{d}_{16}$ is obtained 
from considering various $\pi N$ phase shift analyses as input to analyze the process
$\pi N \to \pi \pi N$.
\item
We have calculated the NN phase shifts in the chiral limit. 
In the $^1S_0$ and $^3S_1-{^3D}_1$ channels we obtain predictions with an error  
governed by the uncertainty in the values of  $\bar D_{\rm S,T}$. 
Both S--wave phase shifts
look qualitatively similar to the physically realized case, although the $^1S_0$ phase 
shift is about $50 - 70 \%$ smaller in magnitude. The $^3P_0$ phase shift is enhanced but no 
bound state appears in that channel, as it is also the case for other P--and 
higher partial waves. Starting from 
D-- and F--waves, the phase shifts nearly reach their asymptotic behaviour caused by the exchange of
a massless pion and grow linearly with momentum, $\delta (k) \sim k$.
\item
According to our analysis, the deuteron is significantly stronger bound in the chiral limit.
The binding energy is close to the natural value of $\sim 10$ MeV one expects to arise in QCD. 
There is no bound state in the $^1S_0$ channel in the chiral limit.
\item
In the chiral limit, the
S--wave scattering lengths take smaller (in magnitude) and more natural values as
compared to the real world.
Furthermore, for pion masses significantly larger than the physical one we found 
negative  values of natural size for the $^3S_1$ scattering length, which seem to 
be consistent with the lattice calculation
\cite{Fukug95}. 
In the $^1S_0$ channel the uncertainty due to the unknown LEC $\bar D_{^1S_0}$
is much larger than in the triplet channel and 
we were not able to predict $a^{^1S_0}$ for 
large values of the pion mass. 
The scattering length takes both positive and negative values with the magnitude 
changing from being natural to infinitely large.
We also consider inverse scattering lengths, which are more 
suitable for chiral extrapolation and for comparison with the lattice calculation.
\item
We have shown that it is possible that Wigner symmetry is also exact in the chiral
limit and that its breaking observed in nature is entirely due to the quark mass terms.
This would explain why the LECs related to Wigner symmetry breaking turn out to be
so much smaller than all other LECs parameterizing the short--distance part of the
NN interaction.
\end{itemize}

To conclude, although we did not find dramatic changes in the properties of the NN systems 
in the chiral limit, the physically realized value of $\tilde M_\pi$ turns out to 
be a quite specific one 
in the sense that it leads to unnaturally large scattering lengths in both 
S--waves and consequently to the small deuteron binding energy. In fact, nature seems to be
more simple in a world with massless quarks since nuclear binding would be of its expected
size and the S--wave scattering lengths would be natural. Thus, no fine tuning would be
needed and the nuclear force problem would be amenable to a much simpler treatment than 
required in the real world. Of course, our investigation can not answer the central
question why nature chooses such a fine-tuned scenario for the nuclear forces, but at least
it is comforting to know that she seems to be more kind in the chiral limit.

\section*{Acknowledgments}
\def\theequation{\arabic{section}.\arabic{equation}}
\setcounter{equation}{0}
\label{sec:ackn}
We would like to thank Silas Beane and Martin Savage for useful discussions
and Nadia Fettes for helpful comments.
E.E.~would like to thank for the hospitality of the Science Center in Benasque,
where a part of this work has been done. This work has been partially supported by the 
Deutsche Forschungsgemeinschaft (E.E.).


\begin{thebibliography}{99}

\bibitem{wein} S.~Weinberg, Nucl. Phys. B363 (1991) 3. \vs
\bibitem{EGM1} E.~Epelbaoum, W.~Gl\"ockle and  Ulf--G.~Mei\3ner, 
Nucl. Phys. A637 (1998) 107. \vs
\bibitem{EGM2} E.~Epelbaum, W.~Gl\"ockle and  Ulf--G.~Mei\3ner, 
Nucl. Phys.  A671 (2000) 295. \vs
\bibitem{3Nno3NF} E.~Epelbaum, et al., arXiv:nucl-th/0201064, accepted for
publication in Eur.~Phys.~J.~A.\vs
\bibitem{new3N} E.~Epelbaum, et al., in preparation.\vs
\bibitem{GL84} J.~Gasser, H.~Leutwyler, Ann. Phys. 158 (1984) 142. \vs
\bibitem{twoloop} 
S.~Bellucci, J.~Gasser and M.~E.~Sainio,
Nucl.\ Phys.\ B423 (1994) 80 [Erratum-ibid.\ B431 (1994) 413];
%[arXiv:hep-ph/9401206];
E.~Golowich and J.~Kambor, Nucl.\ Phys.\ B447 (1995) 373;
%[arXiv:hep-ph/9501318];
J.~Bijnens et al., Phys.\ Lett.\ B374 (1996) 210;
%[arXiv:hep-ph/9511397];
H.~W.~Fearing and S.~Scherer, Phys.\ Rev.\ D53 (1996) 315;
%[arXiv:hep-ph/9408346];
P.~Post and K.~Schilcher, Phys.\ Rev.\ Lett.\   79 (1997) 4088;
%[arXiv:hep-ph/9701422];
J.~Bijnens, G.~Colangelo and G.~Ecker, JHEP  9902 (1999) 020;
%[arXiv:hep-ph/9902437];
J.~Bijnens, G.~Colangelo and G.~Ecker,
Ann. Phys.\  280 (2000) 100.\vs
%[arXiv:hep-ph/9907333].\vs
\bibitem{sanspion}
S.~R.~Beane and M.~J.~Savage, Nucl. Phys. A694  (2001) 511.\vs
\bibitem{uzan}
J.~P.~Uzan, arXiv:hep-ph/0205340.\vs
\bibitem{Bul97} A.~Bulgac, G.A.~Miller, and M.~Strikman, Phys. Rev. C56 (1997) 3307. \vs
\bibitem{Beane01} S.R.~Beane, P.F.~Bedaque, M.J.~Savage, and U.~van Kolck, 
Nucl. Phys. A700 (2002) 377.\vs
\bibitem{BeaneLimC} S.R.~Beane, et al., Phys. Rev. A64 (2001) 042103.\vs
\bibitem{Fukug95} M.~Fukugita, et al., Phys. Rev. D52 (1995) 3003.\vs
\bibitem{BeaneNEW} S.R.~Beane and M.J.~Savage, hep-ph/0206113.\vs
\bibitem{Phil59} R.J.N.~Phillips, Reports on Progress in Physics XXII (1959) 562.\vs
\bibitem{KBW} N.~Kaiser, R.~Brockmann, and W.~Weise, Nucl. Phys. A625 (1997) 758.\vs
\bibitem{allKaiser} N.~Kaiser, Phys. Rev. C61 (2000) 014003, Phys. Rev. C62 (2000) 024001,
Phys. Rev. C63 (2001) 044010, Phys. Rev. C64 (2001) 057001, Phys. Rev. C65 (2002) 017001.\vs
\bibitem{Ta45} I.~Tamm, J. Phys., U.S.S.R., 9 (1945) 449.\vs
\bibitem{Dan50} S.M.~Dancoff, Phys. Rev. 78 (1950) 382.\vs
\bibitem{Fuk54} N.~Fukuda, K.~Sawada, and M.~Taketani,  
Progr. Theor. Phys., Japan, 12 (1954) 156.\vs 
\bibitem{Okubo54} S.~Okubo, Progr. Theor. Phys., Japan, 12 (1954) 603.\vs 
\bibitem{BKMprog} V.~Bernard, N.~Kaiser, and Ulf--G.~Mei\3ner, 
Int. J. Mod. Phys. E4 (1995) 193. \vs
\bibitem{Eck96} G.~Ecker, M.~Moj\v zi\v s, Phys. Lett. B365 (1996) 312.\vs   
\bibitem{Fet98} N.~Fettes, Ulf--G.~Mei\3ner, S.~Steininger, Phys. Lett. B451 (1999) 233.\vs
\bibitem{Fet00} N.~Fettes, Ulf--G.~Mei\3ner, Nucl. Phys. A676 (2000) 311.\vs
\bibitem{FMMS}
N.~Fettes, Ulf-G.~Mei{\ss}ner, M.~Moj\v zi\v s and S.~Steininger,
Annals Phys.\  283 (2000) 273
[Erratum-ibid.\   288 (2001) 249].\vs
\bibitem{Kr99} A.~Kr\"uger, and W.~Gl\"ockle, Phys. Rev. C60 (1999) 024004.\vs
\bibitem{Kr00} A.~Kr\"uger, doctoral thesis, Bochum 2000, unpublished.\vs
\bibitem{Ger71} I.S.~Gerstein, R.~Jackiw, B.W.~Lee, and S.~Weinberg, 
Phys. Rev. D3 (1971) 2486.\vs
\bibitem{NPSA}  V.G.J.~Stoks, R.A.M.~Klomp, C.P.F.~Terheggen and J.J.~de Swart, Phys. Rev. 
C49 (1994) 2950.\vs
\bibitem{GML54} M.~Gell--Mann, and F.E.~Low, Phys. Rev. 95 (1954) 1300.\vs
\bibitem{Chew54} G.F.~Chew, Phys. Rev. 94 (1954) 1748.\vs
\bibitem{Wick55} G.C.~Wick, Rev. Mod. Phys. 27 (1955) 339.\vs
\bibitem{EM}G.~Ecker and M.~Moj\v zi\v s, Phys.\ Lett.\ B410 (1997) 266.\vs
\bibitem{FMSwf}
S.~Steininger, Ulf-G.~Mei{\ss}ner and N.~Fettes, JHEP  9809 (1998) 008.\vs
\bibitem{HD98} J.F.~Donoghue, and B.R.~Holstein, Phys. Lett. B436 (1998) 331.\vs
\bibitem{Burga98}  J.F.~Donoghue, B.R.~Holstein, and B.~Borasoy, Phys.Rev. D59 (1999) 036002.\vs
\bibitem{Fet00_2} N.~Fettes, V.~Bernard, and Ulf--G.~Mei\3ner, Nucl. Phys. A669 (2000) 269.\vs
\bibitem{FetTH} N.~Fettes, doctoral thesis, 
published in {\it Berichte des Forschungszentrum J\"ulich}, No. 3814 (2000); 
N.~Fettes, private communication.\vs
\bibitem{Em98} E.~Matsinos, hep-ph/9807395.\vs
\bibitem{Ka85} R.~Koch, Nucl. Phys. A448 (1986) 707.\vs
\bibitem{Sp98} SAID on--line program, R.A.~Arndt, R.L.~Workman et al., see website 
http://gwdac.phys.gwu.edu/.\vs
\bibitem{EGMres}  E.~Epelbaum, Ulf--G.~Mei\3ner, W.~Gl\"ockle, and Ch.~Elster, 
Phys.Rev. C65 (2002) 044001.\vs
\bibitem{BS1} S.R.~Beane and M.J.~Savage, Phys. Lett. B535 (2002) 177.\vs
\bibitem{BS2} S.R.~Beane and M.J.~Savage, hep-lat/0203003.\vs
\end{thebibliography}
\end{document}